\documentclass[12pt]{article}

\usepackage{amsmath}
\usepackage{amssymb}
\usepackage{amsthm}

\usepackage{upref}

\newtheorem{thm}{Theorem}[section]
\newtheorem{cor}{Corollary}[section]

\newcommand\reals{\mathbb R}

\newcommand\intt{{\rm INT}}
\newcommand\path{{\rm PATH}}

\newcommand\ewor{e^{{\rm wor}}}
\newcommand\eran{e^{{\rm ran}}}
\newcommand\eqs{e^{{\rm qua-std}}}
\newcommand\eqr{e^{{\rm qua-ran}}}
\newcommand\cques{{\rm comp}^{{\rm que-std}}}
\newcommand\cqubs{{\rm comp}^{{\rm qub-std}}}
\newcommand\cquer{{\rm comp}^{{\rm que-ran}}}
\newcommand\cqubr{{\rm comp}^{{\rm qub-ran}}}
\newcommand\cw{{\rm comp}^{{\rm inf-wor}}}
\newcommand\crr{{\rm comp}^{{\rm inf-ran}}}

\def\a{{\alpha}}
\def\e{{\varepsilon}}
\def\l{{\lambda}}

\title{\bf The Quantum Setting \\ with  Randomized  Queries\\
for Continuous Problems}  
\author{H. Wo\'zniakowski\footnote{
This research has been supported in part by 
the Defense Advanced Research
Projects Agency (DARPA) and the National Science Foundation.} \\
$\qquad$\\
{\small Department of Computer Science,} 
{\small Columbia University, New York, USA, and} \\
{\small Institute of Applied Mathematics and Mechanics,} 
{\small University of Warsaw, Poland}}
\date{\today}

\begin{document}
\setcounter{page}{1}
\maketitle
\begin{abstract} 
The standard setting of quantum computation 
for continuous problems 
uses deterministic queries and the
only source of randomness for quantum algorithms 
is through measurement. Without loss 
of generality we may consider quantum algorithms which 
use only one measurement. This
setting is related to the worst case setting on a classical computer
in the sense that the number of qubits 
needed to solve a continuous problem 
must be at least equal to the 
logarithm of the worst case information complexity of this problem.
Since the number of qubits must be finite, 
we cannot solve continuous problems on a quantum computer
with infinite worst case information complexity.
This can even happen for continuous problems with small
randomized complexity on a classical computer. 
A simple example  
is  integration of bounded continuous functions.

To overcome this bad property that limits the power of quantum
computation for continuous problems, 
we study the quantum setting in which {\em randomized}
queries are allowed. This type of query
is used in Shor's algorithm. The quantum setting 
with randomized queries
is related to the randomized classical setting in the sense
that the number of qubits needed to solve a continuous problem 
must be at least
equal to the logarithm of the randomized information 
complexity of this problem.
Hence, there is also a limit to the power of the quantum setting
with randomized queries 
since we cannot solve continuous problems with infinite
randomized information complexity. An example 
is approximation of bounded continuous functions. 

We study the quantum setting with randomized queries
for a number of problems
in terms of the query and qubit complexities defined as the minimal
number of queries/qubits needed to solve the problem to within~$\e$
by a quantum algorithm. We prove that for 
path integration we have an {\em exponential} improvement for
the qubit complexity over the quantum setting with deterministic queries.
\end{abstract}
\section{Introduction}
One of the challenging problems of computational theory is the study of
the power of quantum computation. By now there seems to be agreement 
about the standard setting of quantum computation, see 
\cite{beals,heinrich,nielsen}. This 
setting describes quantum computation as a sequence of 
unitary $2^k\times 2^k$ matrices acting on an
initial state followed by a measurement. Here, $k$
denotes the number of qubits. The unitary matrices in quantum
computation are represented by elementary quantum gates, and one of
them may represent a {\em query}
which depends on the problem we want to solve.
For continuous problems, queries are {\em deterministic} 
and depend on function values. 
Quantum algorithms may have many measurements but it is known that
without loss of generality it is enough to consider quantum algorithms with
only one measurement, see Remarks 1 and 2 and papers cited there. 
 
In what follows it is important to stress the difference between the
cost of an algorithm for solving a given problem and the computational
complexity of this problem. The computational complexity (for
brevity, the complexity) is the {\it minimal} computational resources
needed to solve the problem. Examples of computational resources which
have been studied include memory, time, and communications on a
classical computer, and qubits, quantum gates and queries on a quantum
computer.

We study quantum computation for continuous problems
which are usually defined on spaces of functions. 
The quantum complexity of continuous problems 
has been studied in many papers and 
different queries, such as bit, phase and power queries, 
have been analyzed in the literature, see e.g., 
\cite{Bessen,heinrich,H03,H04a,H04b,jaksch,
Kacewicz,nielsen,novak,N01,PW04,TW02}.
In this paper we study bit queries, although some results hold for
more general queries. What is important for our study is that a query
for a continuous problem depends on at most $2^k$ function values 
computed at a priori given {\em deterministic} sample points. 
This means that in the standard quantum setting
for continuous problems, 
quantum algorithms depend on at most $2^k$
function values and the only source of randomness comes through measurement. 

The {\em query} complexity has been the focus of research. It is  
defined as the minimal number of queries needed to
solve a given problem to within~$\e$ by a quantum algorithm.
Since  a critical resource for the foreseeable future is the number of
qubits, we also study the {\em qubit} complexity which is 
defined as the minimal
number of qubits needed to solve a given problem to within $\e$
by a quantum algorithm.

We stress that there could be a trade-off between the query and qubit
complexities since the minimization of queries may lead to a large
number of qubits and vice versa. We do not know of such
trade-offs for continuous problems studied so far. 
It is unknown if such trade-offs occur for some continuous problems.  

To compute the quantum speedup one needs to know 
the worst case and randomized complexities on a classical computer.
For continuous problems, the worst case and randomized classical
complexities 
have been thoroughly studied in information-based complexity,
see \cite{heinrich,novak,plaskota,traub,TW98,wer}. For our purpose, we need 
the concept of (non-adaptive) information complexity 
which is defined as the minimal
number of function values needed to solve the problem to within~$\e$. 
We included two short sections on these classical settings 
to the extent needed in the rest of the paper. 

Our first technical result is a relation between the standard quantum
setting and the worst case classical setting. Namely, it is relatively easy
to see that since quantum algorithms 
are based on at most $2^k$ function values, they can not have a quantum 
error smaller than the worst case error
of a classical algorithm based on these $2^k$ function values.
This analogy is not complete since in the worst case setting we use 
deterministic algorithms whereas a quantum algorithm has a random
element through measurement. Nevertheless, it is possible to show
that the qubit complexity is bounded from below by the
logarithm\footnote{All logarithms in this paper are base $2$.}
of the worst case information complexity of the problem 
which we want to solve to
within $2\e$. We will show that this extra factor $2$ 
takes care of randomness of
quantum algorithms\footnote{As indicated in the proof of Theorem
3.1 the extra factor $2$ can be often omitted.}. 
Since the worst case information complexity usually 
goes to infinity as $\e$ tends to zero, the number of
qubits must also increase to infinity although at a much slower rate
due to the presence of the logarithm. 

We also show that the qubit complexity is bounded from below by the
(Kolmogorov) $\e$-entropy of the solution set. Hence, problems with large entropy of the
solution set require a large number of qubits. 

When the worst case information complexity or the entropy 
of the solution set is infinite then a finite number of qubits 
is not enough and the problem is unsolvable in the standard quantum setting. 
This can even happen for problems 
for which the randomized classical complexity is small.
An example of such a problem is multivariate integration of continuous
$d$-variate functions defined on, say, $[0,1]^d$, whose absolute values are
bounded by~$1$. It is known that in this case the worst case
information complexity is infinite but the Monte Carlo is optimal and 
the randomized information (as well as the total) complexity 
is roughly $\e^{-2}$ independent of~$d$.

Why can we solve this problem in the classical randomized setting and not in
the standard quantum setting? The reason is that in the randomized
setting we use function values at {\em randomized} points and potentially we
can compute the function value at any point whereas in the standard
quantum setting we use function values at {\em deterministic} points.
The number of these points can be enormous, up to $2^k$.
But if we take a continuous function which vanishes at these $2^k$ 
points then we are unable to detect whether this function is zero or
perhaps takes values equal to $1$ or~$-1$ at all points except an
arbitrarily small neighborhoods of points at which it vanishes.
The true solution may be zero or arbitrarily close to $1$ 
or to~$-1$. That is why any quantum algorithm in the standard quantum
setting must fail.

This negative result is our point of departure. To overcome this bad
property of quantum algorithms and to enlarge the power of quantum
computation we propose a small modification of the standard quantum
setting by allowing the use of {\em randomized} queries and {\em randomized}
unitary matrices. The other assumptions are kept intact.
We will call this modification as
the {\em quantum setting with randomized queries}
and refer sometimes to the standard quantum setting as
the {\em quantum setting with deterministic queries}.

In fact, the idea of using randomized queries is not new. 
A particular kind of randomized query is 
used in Shor's algorithm for factoring of a (large) integer $N$,
see \cite{shor} and also \cite{nielsen}. The
essential part of Shor's algorithm is order finding which is solved by
the query
$$
Q_x|j\rangle\,=\,|j\,x\ {\rm mod}\, N \,\rangle
$$
for $j=0,1,\dots,2^{\lceil \log\,N\rceil}-1\}$ with a {\it random} $x$
from $\{2,3,\dots,N-1\}$. 

The use of randomized queries for continuous problems was also
suggested  in \cite{N01} for integration of non-smooth functions. 

The quantum setting with randomized queries is the same as the standard
quantum setting with the one important exception that queries as well as
all unitary matrices used by a quantum algorithm 
may now depend on a random element. Hence, we have now two sources 
of randomness: one affecting unitary matrices and the other affecting 
measurement. 

The quantum setting with randomized queries 
is an extension of the standard quantum
setting. Indeed, if one always selects the same 
unitary matrices including the query, then we have exactly the
standard quantum setting.
Obviously, the use of randomized unitary matrices and randomized 
queries offers a possibility of much more efficient quantum computation.
For some continuous problems, this extension is necessary. For
example, we will show that multivariate integration of bounded 
and continuous functions,
which cannot be solved in the standard quantum setting, is solvable in
the  quantum setting with randomized queries by a quantum algorithm that uses 
of order $\e^{-1}$ queries and $\log\,\e^{-1}$ qubits. Hence, we have 
a quadratic speedup over the randomized setting on a classical computer.

We now comment on the error criteria used in the 
quantum settings with deterministic and randomized queries. 
In the standard quantum setting, i.e., in the quantum setting with 
deterministic queries, two error criteria are studied: 
\begin{itemize}
\item the first error criterion is defined by taking
 the average performance of a quantum algorithm 
 with respect to measurements for a worst
function from the given class, 
\item the second error criterion is defined by taking 
 the worst case performance of a quantum algorithm 
 with respect to measurements on a
 set of measure $1-\delta$ for a worst function from the given class. 
\end{itemize}
The same error criteria are used in the quantum setting with
randomized queries. In this case, randomization is richer and we take
the average performance or a set of measure $1-\delta$ 
with respect to
``measurements, randomized queries and randomized unitary
matrices''.
The first error criterion is studied in the main body of the
paper whereas the second one is studied in the appendix.

We define the {\em randomized} (bit) query and qubit complexities 
analogously to the randomized setting on a classical
computer. The randomized query complexity is defined as the minimum of the
average number of randomized queries needed to solve the problem
to within~$\e$ by a quantum algorithm.
By ``to within~$\e$'', we now mean that the error
of a quantum algorithm is at most $\e$ which is defined by
taking the average performance with respect to all random elements of the
quantum algorithm for a worst function from the given class. 

The randomized qubit complexity is defined analogously as
the minimal number of qubits for which there is a quantum
algorithm whose error is at most
$\e$. We stress that we assume 
the number of qubits is fixed and does not vary
during quantum computation. 
This is probably a reasonable assumption 
from a practical point of view since a quantum computer with a random
number of qubits seems too much to be expected in the near future.
Nevertheless, from a purely theoretical point of view 
it would be interesting to study also the quantum setting with
randomized queries and with random number of qubits and try to 
minimize the average number
of qubits needed to solve the problem. 

It is not surprising that the quantum setting with randomized queries 
is related to the randomized setting on a classical computer 
in the sense that the randomized qubit
complexity is bounded from below by the logarithm of the randomized
(non-adaptive) information complexity on a classical computer. 
If the randomized information complexity of a problem is
infinite, the problem cannot be solved 
in the quantum setting with randomized queries. This happens, for
example, for approximation of bounded continuous functions.
Of course, the class of problems with infinite randomized
information complexity
is smaller that the class of problems with infinite worst case
information complexity. 
So we extend the limit of what can be computed by presenting
the quantum setting with randomized queries.

We study the quantum setting with randomized queries 
for a number of problems,
and for some of them we prove an {\em exponential} improvement for
the qubit complexity compared to the standard quantum setting.
This is especially important since, as already mentioned,
the number of qubits is a critical resource for the foreseeable
future. In particular, the exponential improvement holds for path integration.

In this paper, we study real and Boolean summation, 
multivariate integration and path integration. We now briefly state the results
obtained for these problems. 

The real summation problem lies at the core of many continuous
algorithms and plays a major role in the study of continuous problems
in the standard quantum setting. The same is true in the 
quantum setting with randomized queries. 
It is known, see e.g., \cite{heinrich,novak}, that 
the real summation problem can be reduced to Boolean summation.
That is why it is enough to present in detail results for only the latter
problem in which we want to approximate
$$
{\cal B}_N(f)\,=\,\frac1N\,\sum_{j=0}^{N-1}f(j)
$$
for a Boolean function $f:\{0,1,\dots,N-1\}\to\{0,1\}$. Here $N$ is a
large integer which can
be assumed to be a power of $2$. We want to compute ${\cal B}_N(f)$ to
within $\e$. Without loss of generality we may consider
$\e^{-1}\ll N$. We now present the orders of the query and qubit 
complexities in the quantum settings with deterministic and randomized
queries.
\vskip 1pc
\begin{center}
\begin{tabular}{|c|c|c|}
\hline
 &            &          \\
& {Deterministic Queries} & {Randomized Queries}\\
 &            &          \\
\hline
 &            &          \\
Query Complexity
 &  $\e^{-1}$ & $\e^{-1}$ \\
 &            &           \\
\hline
 &            &          \\
Qubit Complexity & $\log\,N$ & $\log\,\e^{-1}$\\
 &            &          \\
\hline
\end{tabular}
 \end{center}
\begin{center}
Figure 1: Boolean Summation
\end{center}
\vskip 1pc
We stress that the minimal numbers of queries and qubits in a given
setting are obtained by essentially the
same quantum algorithm. In the quantum setting
with deterministic queries, this is the Boolean 
summation algorithm of \cite{brassard} with seven repetitions as proved in
\cite{hkw}. In the quantum setting with randomized queries,  
we first approximate the Boolean mean ${\cal B}_N(f)$ by
the Monte Carlo algorithm,
$$
\mbox{MC}_m(f,\omega)\,=\,\frac1m\,\sum_{j=1}^mf(\omega_j),
$$
with $m$ of order $\e^{-2}$ and with independent and uniformly distributed
$\omega_j$ from $\{0,1,\dots,N-1\}$, and then use 
the Boolean summation algorithm  with seven repetitions to 
approximate $\mbox{MC}_m(f,\omega)$. 
It is interesting to notice that this algorithm uses randomized
queries but the remaining unitary matrices are deterministic. 
This leads to the
corresponding upper bounds.  
The lower bound proof of the query complexity
in the standard case follows from \cite{nayak}.
For the quantum setting with randomized queries, 
we use the known fact that the
randomized errors of quantum algorithms are no smaller than the average case errors
with respect to Boolean functions. The latter problem with an
appropriate measure on Boolean functions was solved in \cite{Papa}.
The lower bound proof of the qubit complexity is from the relation to
the randomized information complexity. 

We stress that we have the same order of 
query complexities in both cases.
However, by allowing randomized queries
we have an essential improvement in the number of qubits
for solving the Boolean summation problem. 

For the real summation problem, we want to approximate
$$
\mbox{SUM}_N(f)\,=\,\frac1N\,\sum_{j=0}^{N-1}f(j),
$$
where $f:\{0,1,\dots,N-1\}\to [0,1]$ may now take real values. For
completeness, in Section~5 we show how the real summation problem
may be reduced to the Boolean summation problem.
In Corollary 5.1 we show that
the results presented in Figure 1 also hold for the real summation problem.  

We now turn to multivariate integration for functions which are
$r$ times differentiable and uniformly bounded. For $r=0$, 
the query and qubit complexities are infinity in 
the quantum setting with deterministic queries. For the quantum
setting with randomized queries, 
they are finite and their orders are given in the
following table.  
\vskip 1pc
\begin{center}
\begin{tabular}{|c|c|c|}
\hline
 &            &          \\
& {Deterministic Queries} & {Randomized Queries}\\
 &            &          \\
\hline
 &            &          \\
Query Complexity
 &  $\infty$ & $\e^{-1}$ \\
 &            &           \\
\hline
 &            &          \\
Qubit Complexity 
 & $\infty$ & $\log\,\e^{-1}$\\
 &            &          \\
\hline
\end{tabular}
 \end{center}
\begin{center}
Figure 2: Multivariate Integration for $r=0$
\end{center}

Hence, in this case the improvement of the quantum setting with
randomized queries over the standard quantum setting is infinite.

We now assume that $r\ge1$. Hence, functions are now at least once 
differentiable. In this case the orders of the 
query and qubits complexities are the same in both cases. 
\vskip 1pc
\begin{center}
\begin{tabular}{|c|c|c|}
\hline
 &            &          \\
& {Deterministic Queries} & {Randomized Queries}\\
 &            &          \\
\hline
 &            &          \\
Query Complexity
 &  $\e^{-1/(1+r/d)}$ & $\e^{-1/(1+r/d)}$ \\
 &            &           \\
\hline
 &            &          \\
Qubit Complexity 
 & $\log\,\e^{-1}$ & $\log\,\e^{-1}$\\
 &            &          \\
\hline
\end{tabular}
 \end{center}
\begin{center}
Figure 3: Multivariate Integration for $r\ge1$
\end{center}

The query complexity in the standard quantum setting is due to
\cite{N01}. The randomized query complexity   
has the same order since Boolean and real summation require roughly the
same queries in both settings. The qubit complexities are of the same
order since the logarithms of the worst case and randomized
information complexities of multivariate integration are both proportional to
$\log\,\e^{-1}$. 

Finally, we consider a specific case of path integration
studied in \cite{TW02}. The orders of query and qubit complexities are
presented in the following table.  
\begin{center}
\begin{tabular}{|c|c|c|}
\hline
 &            &          \\
& {Deterministic Queries} & {Randomized Queries}\\
 &            &          \\
\hline
 &            &          \\
Query Complexity
 &  $\e^{-1+o(1)}$ & $\e^{-1+o(1)}$ \\
 &            &           \\
\hline
 &            &          \\
Qubit Complexity 
 & $\e^{-2}\log\,\e^{-1}$ & $\log\,\e^{-1}$\\
 &            &          \\
\hline
\end{tabular}
 \end{center}
\begin{center}
Figure 4: Path Integration
\end{center}

We thus have the same orders of query  complexities and an exponential
improvement in the number of qubits. 

\vskip 1pc
We stress that in the randomized classical setting and in the quantum 
setting with randomized queries
we permit the use of random elements from a set $\Omega$
whose cardinality may be infinite and distribution of
points from $\Omega$ may be arbitrary. For example,
the classical Monte Carlo with $n$ random points 
for integration of $d$-variate functions 
defined over, say, $[0,1]^d$, uses $\Omega=[0,1]^{\,dn}$
and uniform distribution. Alternatively, it is possible, also  
for classical computers, to use a restricted form of randomization
based on, for example, random bits or a finite set $\Omega$.
This obviously restricts the class
of randomized algorithms and it is not clear if positive results
for unrestrictive randomization are still true for the restricted 
case. There is a very interesting stream of work,
see \cite{HNP,Pfeiffer}, studying the minimal number of random bits needed
for the solution of continuous problems on a classical computer.
There are also general results in \cite{Stefannew}
showing that as long as $\Omega$  is finite then
the classical randomized setting
is (roughly) equivalent to the standard quantum setting at the
expense of adding additional qubits. 
However, if the cardinality of $\Omega$ goes to infinity then the
additional number of qubits also goes to infinity. That is why,
the standard quantum setting is {\it not} 
equivalent to the classical randomized setting without a restriction on 
$\Omega$.

We hope that the quantum setting with (restricted) randomized queries 
will be studied for general continuous problems. 
It would be especially interesting
to characterize continuous problems for which this setting
offers an exponential improvement in the number of queries
and/or qubits over the standard quantum setting. 

\section{Continuous Problems}
The computational complexity of approximate solutions of continuous
problems has been studied in information-based complexity, see e.g., 
\cite{heinrich,novak,plaskota,traub,TW98,wer}. We present a brief
outline of this theory in the worst case, randomized and quantum
settings to the extent needed for this paper.

Let $F$ be a non-empty subset of a linear space of $d$-variate
functions $f:D_d\to\reals$ with $D_d\subset\reals^d$. Let $G$ be a
normed space with its norm denoted by $\|\cdot\|$. Consider a (linear
or non-linear) operator
$$
S\,:\ F\ \to\ G.
$$
Our goal is to compute $S(f)$ to within $\e$ for $f\in F$.
\vskip 1pc
\noindent {\bf Example\,: Multivariate Integration}\ 
\newline
We illustrate the concepts of this paper by an
example of multivariate integration of smooth functions. 
Let $C^r([0,1]^d)$ denote the class of real functions defined on the
$d$-dimensional unit cube, $f:[0,1]^d\to\reals$, all of whose partial
derivatives up to order $r$ exist and are continuous. That is, for a
multi-index $\a=[\a_1,\a_2,\dots,\a_d]$ with non-negative integers
$\a_j$ and with $|\a|:=\a_1+\a_2+\cdots+\a_d\,\le\,r$ we know that
$$
D^{\,\a}f\,=\,\frac{\partial^{\,|\a|}}{\partial t_1^{\,\a_1}\partial
    t_2^{\,\a_2}\cdots \partial t_d^{\,\a_d}}\,f
$$
exists and is continuous. The norm in $C^r([0,1]^d)$ is defined as
$$
\|f\|_r\,=\,\max_{\a:\,|\a|\le r}\ \max_{t\in[0,1]^d}\,|D^{\a}f\,(t)|. 
$$
Then we set 
$$
F\,=\,F_{d,r}\,=\,\{\,f\in C^r([0,1]^d):\ \|f\|_r\,\le\,1\,\}
$$
as the unit ball of $C^r([0,1]^d)$, and $G=\reals$. 

The multivariate integration problem  $S=\intt_{d,r}\,:\,C^r([0,1]^d)\,\to\,\reals$ is defined by
$$
\intt_{d,r}(f)\,=\,\int_{[0,1]^d}f(t)\,dt.
$$
This is an example of a linear problem since the operator $\intt_{d,r}$ 
depends linearly on $f$. \ \qed
\vskip 1pc
The approximate computation of $S(f)$ can be done as follows. 
First of all, we specify how information about the function $f$ is used
by algorithms. We assume that we can compute finitely many function
values\footnote{More general information given by arbitrary linear
functionals on $f$ has also been extensively studied in information-based
complexity, see e.g., \cite{traub}}
$f(t)$ for some sample points $t$ from $D_d$.
That is, any algorithm may use $f(t_1),f(t_2),\dots,f(t_n)$ for some
$n$ and $t_j$. We stress that the choice of the sample points
$t_j$ may be {\it adaptive}, i.e., $t_1$ is given a priori, whereas 
$t_j$ may depend on the already computed values
$f(t_1),f(t_2),\dots,f(t_{j-1})$ for $j=2,3,\dots,n$. The number $n$
can also be chosen adaptively. 
The sample points may be chosen deterministically or randomly
depending on the setting, for details see \cite{traub}. 
The specific form of an algorithm  
also depends on the setting in which we define the error.
We first present two settings for classical computers and then turn
our attention to the quantum setting with deterministic
and randomized queries. 

\subsection{Classical Computers:\ Worst Case Setting}

In the worst case setting, we assume that sample points as well as
algorithms are {\it deterministic}. That is, an algorithm that uses
$n$ function values has the form
\begin{equation}\label{algworst}
A_n(f)\,=\,\phi\left(f(t_1),f(t_2),\dots,f(t_n)\right)
\end{equation}
for some mapping $\phi:\reals^n\to G$. If the sample points 
are given a priori and are the same for all $f$ from $F$, then
$A_n$ uses {\it non-adaptive} information. Otherwise, it uses
{\it adaptive} information. 

The worst case error of the
algorithm $A_n$ is given by its worst case performance
with respect to $f$,
$$
\ewor(A_n)\,=\,\sup_{f\in F}\|S(f)-A_n(f)\|.
$$ 
\vskip .5pc
\noindent {\bf Example\,: Multivariate Integration (continued)}\
\newline
A typical choice of an algorithm for multivariate integration is
a linear algorithm, sometimes called a quadrature or cubature,
$$
A_n(f)\,=\,\sum_{j=1}^na_{j}f(t_{j})
$$
for some $a_{j}\in \reals$ and $t_{j}\in[0,1]^d$.

For $r\ge1$, it was proven by Bakhvalov already in 1959, see \cite{Bakh} and
also \cite{novak,traub}, that the minimal worst case error of
algorithms using $n$ function values is proportional to $n^{-r/d}$.
Furthermore, the error of order $n^{-r/d}$ can be achieved by a linear
algorithm using non-adaptive information.  
Hence, if we want to guarantee that  $\ewor(A_n)\le\e$, then $n$
has to be of order $\e^{-d/r}$, and is exponential in $d$. This is
called the {\it curse of dimensionality} meaning that multivariate
integration is {\it intractable} in the worst case setting
if $d$ is much larger that~$r$. The case $r=0$ will be considered
later. 

For large $d$, a popular choice of 
$a_j$ is $n^{-1}$ which leads to QMC (quasi-Monte Carlo) 
algorithms. The sample points are often chosen as low discrepancy
points, lattice or shifted lattice points, see
\cite{Nied,SJ}. For some spaces other 
than $C^r([0,1]^d)$ the error behavior of such algorithms is only
polynomial in $d$ or even independent of $d$ and tends to zero as a
positive power of $n^{-1}$. This is an active research area
of information-based complexity dealing with high dimensional
problems; the reader may consult 
\cite{NW} for a survey. \ \qed

\subsection{Classical Computers: Randomized Setting}

In the randomized setting, we allow randomized choices of sample
points as well as algorithms. That is, we have a probability space of
elements $\omega$ from some  set $\Omega$ which are distributed
according to some probability measure $\rho$ on $\Omega$,
$\rho(\Omega)=1$. Algorithms using $n$ function values on the average 
have now the form
\begin{equation}\label{algrand}
A_n(f;\omega)\,=\,\phi_{\omega}\left(f(t_{1,\omega}),f(t_{2,\omega}),\dots,
f(t_{n_\omega,\omega})\right),
\end{equation}
where $t_{j,\omega}$ are randomized sample points from $D_d$, and
$\phi_{\omega}$ is a randomized mapping from $\reals^{n_\omega}$ to
$G$. Here, $n_\omega$ is the randomized number of sample points and
its average is $n$, i.e., 
$$
n\,=\,\int_{\Omega}n_\omega\,\rho(d\omega).
$$ 
We stress that the sample points $t_{j,\omega}$ as well as $n_\omega$
can be chosen adaptively as in the worst case setting. This also means that
the probability measure $\rho$ may depend on the function  $f$ through 
its computed function values, see Chapter 10 of \cite{traub} for
details. If the sample points $t_{j,\omega}$
are the same for all $f$ from $F$, then $A_n$ uses {\it non-adaptive}
randomized information, otherwise it uses {\it adaptive} randomized
information. 

The {\em randomized} error of the algorithm $A_n$ is defined by its worst
case performance with respect to $f$ and the average performance with
respect to $\omega$, 
\begin{equation}\label{rand}
\eran(A_n)\,=\,\sup_{f\in F}
\left(\int_{\Omega}\|S(f)-A_n(f,\omega)\|^2\rho(d\omega)\right)^{1/2}.
\end{equation}
Here, we choose to study the average performance in the $L_2$-norm;
however it is also possible to study it in a more general case of the
$L_p$-norms with $p\in[1,\infty)$. 
\vskip .5pc
\noindent {\bf Example\,: Multivariate Integration (continued)}\
\newline
Probably the most popular and widely used randomized algorithm is the
Monte Carlo algorithm
$$
A_n(f,\omega) \,:=\,{{\rm MC}}_n(f,\omega)\,=\,
\frac1n\sum_{j=1}^nf(t_{j,\omega}),
$$
where $t_{j,\omega}$ are independent and uniformly distributed sample
points over $[0,1]^d$. In this case, $\Omega=[0,1]^{dn}$ and $\rho$
is Lebesgue's measure. That is,
$\omega=[\omega_1,\omega_2,\dots,\omega_n]$
with $\omega_j\in [0,1]^d$ and $t_{j,\omega}=\omega_j$. 
We stress that Monte Carlo uses non-adaptive randomized
information with deterministic $n$ and the deterministic 
mapping $\phi_{\omega}=\phi$
given by $\phi(y)=n^{-1}\sum_{j=1}^ny_j$. It is well known that 
$$
\int_{[0,1]^{nd}}\left(
\int_{[0,1]^d}f(t)\,dt-\frac1n\sum_{j=1}^nf(\omega_j)\right)^2\,d\omega_1\cdots
d\omega_d\,=\,
\frac{\int_{[0,1]^d}f^2(t)\,dt\,-\,\left(\int_{[0,1]^d}f(t)\,dt\right)^2}{n}.
$$
Since for $f\in F_{d,r}$ with $r\ge0$, we have $\int_{[0,1]^d}f^2(t)dt\le 1$, then
$$
\eran\left({{\rm MC}}_n\right)\,\le\,n^{-1/2}.
$$
Hence, $\eran({{\rm MC}}_n)\le \e$ for $n=\lceil \e^{-2}\rceil$
and the curse of dimensionality of the worst case setting is broken 
by Monte Carlo in the randomized setting. Bakhvalov also
proved, see \cite{Bakh}, that the minimal randomized error of
algorithms using $n$ function values is of order $n^{-1/2+r/d}$,
and the latter error bound is achieved by a linear algorithm using
non-adaptive information.
This bound also holds if we use $n$ function values on the average
as proven by Novak in \cite{Novakran}. Thus, Monte Carlo
almost minimizes the randomized error if $d$ is much larger than $r$.  
\ \qed 
\vskip 1pc
The errors of randomized algorithms may be defined differently
than (\ref{rand}). This corresponds to the probabilistic 
errors which are related to the quantum setting error commonly used in many
papers. To simplify the presentation of the paper, we deal with the
probabilistic errors in the appendix. 

\subsection{Complexity and Information Complexity}
As already mentioned, we want to compute $S(f)$ to within $\e$.
That is, we are looking for an algorithm $A_n$ whose error 
in the worst case or randomized setting is at most $\e$, 
\begin{equation}\label{error}
e^{{\rm wor/ran}}(A_n)\,\le\,\e.
\end{equation}
We would like to guarantee (\ref{error}) with the {\it minimal cost} of
computing $A_n(f)$. This minimal cost is called the (total)
{\it$\e$-complexity of $S$}, and denoted by ${\rm comp}^{{\rm wor/ran}}(\e,S)$.
The cost of computing $y=A_n(f)$ is defined by counting the cost of $n$
function values plus all operations needed to obtain $y$. 
The abstraction typically used in information-based complexity 
(and in scientific computation) is the real number model of
computation in which we assume we can
perform arithmetic operations and comparisons of real numbers with
unit cost independently of the size of numbers, again see \cite{traub,TW98}
for details. The reader is referred to \cite{traub99}
for the motivation behind the real number model and
comparison with the Turing model of computation.

As we shall see, for quantum computation   
the complexity of $S$ is less relevant than 
the non-adaptive information complexity. 
The latter is defined as the minimal
number $n$ of non-adaptive function values in a given setting needed 
to find an algorithm $A_n^{{\rm nad}}$ with error at most $\e$. More precisely,
in the worst case setting, algorithms $A_n^{{\rm nad}}$ are of the
form (\ref{algworst})  with a priori given sample points $t_j$, 
whereas in the randomized setting, they are of the form (\ref{algrand})
with 
sample points $t_{j,\omega}$ independent of $f$ and depending only on
$\omega$ with fixed $n_\omega=n$. Hence, 
the {\it non-adaptive information complexity} is defined by
\begin{equation}\label{non-adaption}
{\rm comp}^{{\rm inf-wor/ran}}(\e,S)\,=\,
\min\left\{\,n\,:\ \exists\, A_n^{{\rm nad}}\ \mbox{such that\ } e^{{\rm
        wor/ran}}(A_n^{{\rm nad}})\,\le\,\e\,\right\}.
\end{equation}
We stress that the minimum in (\ref{non-adaption}) is taken over {\em
  all} algorithms using non-adaptive information. That is, over all
possible sample points $t_j$ and functions $\phi$ in (\ref{algworst}) in the
worst case setting, and over all probability measures $\rho$, sample
points $t_{j,\omega}$ and functions $\phi_{\omega}$ in (\ref{algrand})
in the randomized setting. 

Surprisingly, for many continuous problems 
the non-adaptive information complexity is practically the same as the total
complexity.  There are, however, continuous problems for which the use
of adaptive information is crucial and 
the non-adaptive information complexity is much larger than 
the total complexity, see \cite{TW80} pp. 165-170. 
There are also continuous problems for which the reverse is true.
That is, non-adaptive information complexity is small but 
it is impossible to combine it in a finite number of operations, 
and therefore the total complexity is
infinite, see \cite{WW}.
\vskip .5pc
\noindent {\bf Example\,: Multivariate Integration (continued)}\
\newline
Bakhvalov's results mean that the total and non-adaptive information complexities of
multivariate integration for the unit ball of $C^r([0,1]^d)$ are 
of order 
\begin{eqnarray*}
\e^{-d/r}&& \mbox{in the worst case setting with $r\ge1$,}\\
\e^{-2/(1+2r/d)}&& \mbox{in the randomized setting with $r\ge0$.}
\end{eqnarray*}
\ \qed 
\section{Quantum Setting with Deterministic Queries}

We describe the standard quantum setting by presenting a general form
of quantum algorithms used in this setting and the definition of their errors. 
Quantum algorithms can be characterized, in particular,
by the number of queries and qubits they use. 
If they use $n$ queries and $k$ qubits, we will denote
them by $A_{n,k}$. Queries and qubits
are deterministic and the only source of randomness is through measurement. 
We stress that quantum algorithms use
{\it non-adaptive} information about the functions $f$. 

The quantum algorithms $A_{n,k}$ are of the following 
form, see \cite{brassard,heinrich,novak}. 
All computations are done on unit vectors in the complex space 
${\cal C}^{2^k}$. 
Here, $k$ denotes the number of qubits.
We assume that the initial state is a unit vector 
$|\psi_0\rangle$ from ${\cal C}^{2^k}$. 
For $f\in F$, the final state 
$|\psi_f\rangle$ is equal to
\begin{equation}\label{quantum}
|\psi_f\rangle\,=\,U_nQ_f\,U_{n-1}Q_f\,\cdots\,U_1Q_f\,U_0|\psi_0\rangle,
\end{equation}
where $U_0,U_1,\dots,U_n$ are $2^k\times 2^k$ unitary matrices which
are independent of $f$. Usually it is required that each $U_j$ is
represented by a relatively small number of elementary quantum gates.
This will not be important for our considerations and we permit the 
use of arbitrary unitary matrices $U_j$. Of course, this makes lower
bounds on the number of needed queries and qubits stronger.  

The query $Q_f$ is also a $2^k\times 2^k$ unitary matrix and depends
on the function $f$. We assume in this paper that $Q_f$ is a {\it bit} 
query, see \cite{heinrich,H03,H04a,H04b,nielsen,N01} 
although results on the qubit complexity also hold for more
general queries such as phase and power queries studied in \cite{Bessen,PW04}.

For a Boolean function $f:\{0,1,\dots,2^m-1\}\to\{0,1\}$, with $k=m+1$,
the bit query is defined by 
$$
Q_f|j\rangle|i\rangle\,=\,|j\rangle|i\oplus f(j)\rangle
$$
for all $i\in \{0,1\}$ and $j\in\{0,1,\dots,2^m-1\}$, and $\oplus$
denotes the addition modulo $2$. 

For real functions $f:D_d\to \reals$, we assume\footnote{For simplicity
  we do not consider ancilla qubits.} that $k=m_1+m_2$, where
$m_1$ qubits are needed to code the arguments of $f$ and $m_2$ qubits 
are used for the values of $f$. The coding is done by two mappings
\begin{eqnarray*}
\tau\,:\,\{0,1,\dots,2^{m_1}-1\}\ &\to& \ D_d,\\
\beta\,:\,f(D_d)\ &\to&\ \{0,1,\dots,2^{m_2}-1\},
\end{eqnarray*}
and the bit query takes the form
$$
Q_f|j\rangle|i\rangle\,=\,|j\rangle|i\oplus \beta(f(\tau(j))\rangle
$$
for all $j\in \{0,1,\dots,2^{m_1}-1\}$ and
$i\in\{0,1,\dots,2^{m_2}-1\}$,
and $\oplus$ now means the addition modulo $2^{m_2}$, 
for more details see \cite{heinrich} . 

We stress that the bit query depends on  at most $2^{m_1}$
function values computed at some non-adaptive points
$t_j=\tau(j)$. Furthermore, although we will not use this fact later,
these function values are usually computed with some
noise due to the finite range of the coding function~$\beta$.
Usually, $\beta(f(t_j))$ is defined as the $m_2$ most significant
bits of $f(t_j)$.   
Obviously, $2^{m_1}\le 2^k$. If $f(D_d)$ is bounded then $m_1$
and $m_2$ are usually of the same order. To simplify further considerations
we will use $2^k$ instead of $2^{m_1}$. 

For our purpose, the most important property of the bit query is that
$Q_f$ depends on at most $2^k$ function values
taken at some a priori given (non-adaptive) {\it deterministic} sample
points $t_j$ from $D_d$,
\begin{equation}\label{qf}  
Q_f\,=\,Q_{f(t_1),f(t_2),\dots,f(t_{2^k})}.
\end{equation}

The results on the qubit complexity will be derived using the
property (\ref{qf}). Therefore they will be valid for {\it all} queries
satisfying (\ref{qf}) which hold, in particular, for 
bit, phase and power queries. 

The bit query $Q_f$ is a {\it deterministic} $2^k\times 2^k$  unitary matrix,
and therefore the final state $|\psi_f\rangle$ is also a deterministic vector
of $2^k$ components which use $n$ times the query $Q_f$  based on 
non-adaptive information consisting of at most $2^k$
function values at some sample points. This means that if we consider
two functions $f_1$ and $f_2$ both from $F$ such that
$f_1(t_j)=f_2(t_j)$ for $j=1,2,\dots,2^k$ then the queries $Q_{f_1}$
and $Q_{f_2}$ are the same, and therefore we obtain the same final
states
$$
|\psi_{f_1}\rangle\,=\,|\psi_{f_2}\rangle\
$$
for both $f_1$ and $f_2$.     

The only source of randomness is
through {\it measurement}. That is, we obtain an index
$j\in\{0,1,\dots,2^k-1\}$ with probability $p_{f,j}$ which depends on the 
final state $|\psi_f\rangle$. As before, $p_{f,j}$ depends on the function
$f$ only through its values $f(t_1),f(t_2),\dots,f(t_{2^k})$,
and $\sum_{j=0}^{2^k-1}p_{f,j}=1$. Knowing 
the index $j$, we compute on a classical computer
\begin{equation}\label{algquas}
A_{n,k}(f,j)\,=\phi(j)
\end{equation}
for some mapping $\phi:\{0,1,\dots,2^k-1\}\to G$. 
The algorithm $A_{n,k}$ is called a {\em quantum algorithm}.

We stress that the
quantum algorithm that uses $k$ qubits takes at most $2^k$
different values independently of the number $n$ of queries used.

\vskip 1pc
\noindent{\bf Remark 1:}
We add that, in principle, we may use hybrid algorithms 
that are combinations of classical 
algorithms using bit operations on classical computers 
and quantum algorithms with many
measurements. However, it is known, see e.g.,
\cite{bernstein,heinrich}, that  such algorithms 
may be rewritten in the form (\ref{quantum}) and (\ref{algquas})
with one measurement at the end of computation by linearly increasing
the number of queries and qubits.
It is also known, see Lemma 1 in \cite{H03}, that we can sample
$\Gamma(f)$ instead of $f$ if $\Gamma(f)$ depends on $\kappa$ function
values of $f$. Then the query $Q_{\Gamma(f)}$ can be simulated by 
a quantum algorithm that uses $2\kappa$ queries on $f$.  
Therefore, without loss of generality we may consider 
only quantum algorithms with one measurement 
of the form (\ref{quantum}) and (\ref{algquas}).  

We stress that this is true if hybrid algorithms use only bit
operations on classical computers. If we use the real number model
of computation then not every algorithm can be written in
the form (\ref{quantum}) and (\ref{algquas}). One reason of this is
that we may have infinitely many outputs in the real number model
of computation which is impossible to obtain in the quantum setting. 
\ \qed
\vskip 1pc
We now discuss the error of a quantum algorithm.
There are at least two natural ways to define the error.
One of them is by taking the worst case performance of a quantum 
algorithm with respect to $f$ and the average case performance 
with respect to the index~$j$.
The other is to take the worst case performance with respect to $f$
and the worst case performance with respect to the index~$j$ modulo a
set of measure $\delta$ for some (usually small) positive $\delta$.
For some problems, when $S$ is a linear functional, it is enough
to take, say, $\delta=3/4$ and increase the probability of success
by running the quantum algorithm a couple of times and by taking the
median as the final result. 

We will study both definitions of the error of a quantum algorithm. 
In the main body of the paper we choose
the first option since it is directly related to the error 
usually studied in the randomized classical setting. 
The other error, which is probably more popular in the quantum
literature, is called the probabilistic error 
and is studied in the appendix. 

Hence, by the error of the quantum algorithm $A_{n,k}$ we mean
\begin{equation}\label{randquant}
\eqs(A_{n,k})\,=\,\sup_{f\in F}\ \left(
\sum_{j=0}^{2^k-1}p_{f,j}\|S(f)-A_{n,k}(f,j)\|^2\right)^{1/2}.
\end{equation}
This concludes the definition of the standard quantum setting
which can be summarized by the general form of a quantum algorithm
(\ref{quantum}) and its error (\ref{randquant}). 

We are interested in finding quantum algorithms with error at most
$\e$. We would like to achieve this goal with the minimal number of
queries and/or qubits. 

By the {\it query complexity in the
standard quantum setting} we mean  
\begin{equation}\label{qques}
\cques(\e,S)\,=\,\min\left\{\,n\,:\ \exists\,A_{n,k}\ \mbox{such that\
  } \eqs(A_{n,k})\,\le\,\e\ \right\}.
\end{equation}
By ``there exists $A_{n,k}$'' we mean a quantum algorithm using $n$
queries and $k$ qubits with a finite $k$ which can be, however,
arbitrarily large. Hence, it may happen that the minimization of the number of queries
will be possible at the expense of the number of qubits. 

By the {\it qubit complexity in the standard quantum setting}
we mean 
\begin{equation}\label{qqubs}
\cqubs(\e,S)\,=\,\min\left\{\,k\,:\ \exists\,A_{n,k}\ \mbox{such that\
  } \eqs(A_{n,k})\,\le\,\e\ \right\}.
\end{equation}
In this case, by ``there exists $A_{n,k}$'' we mean an arbitrary
choice of the number $n$ of queries. Obviously, $k$ in (\ref{qques})
must be at least as large as $\cqubs(\e,S)$, and $n$ in 
(\ref{qqubs}) 
must be at least as large as $\cques(\e,S)$.

Although we do not pursue this point in the paper, it is also
reasonable to minimize both queries and qubits. For example, 
we may want to minimize $k\,n$ or  
$n+\beta\,k$, for a given positive number $\beta$, 
over all quantum algorithms using $n$ queries and $k$
qubits whose quantum error is at most $\e$. Here, if we choose $\beta$
small then our emphasis will be on the number of queries, and if
$\beta$ is large then our emphasis will be on the number of qubits.
\vskip 1pc
\noindent{\bf Remark 2:}\ We stress that query and qubit
complexities are defined by minimizing the number of queries/qubits
needed to solve the problem by a quantum algorithm with one
measurement.

Suppose we have quantum computation which requires the use of 
a sequence of quantum algorithms $A_{n_j,k_j}$ each with one measurement
and uses $n_j$ queries and $k_j$ qubits for $j=1,2,\dots,p$. Then 
the total number of queries is $n=\sum_{j=1}^pn_j$ which
is of the same order when all $A_{n_j,k_j}$ are transformed as 
a quantum algorithm $A$ with one measurement
which uses $O(n)$ queries. 

The situation is, however, different for qubits since to run all
quantum algorithms $A_{n_j,k_j}$ it is enough 
to have $k=\max_{j=1,2,\dots,p}k_j$ qubits
whereas the quantum algorithm $A$ would require of order
$\sum_{j=1}^pk_j$ qubits. Obviously, as long as $p$ does not depend
on $\e$, it does not really matter since 
$k$ must be at least of the same order as the qubit complexity.
If, however, $p$ is large and depends on $\e$, then
the qubit complexity needed for quantum algorithms with one
measurement may be improved by many measurements. 

There is one case for which the size of $p$ does not matter. Namely,
when the qubit complexity is infinite which can happen as we see in
the next section. 

It would be tempting to redefine the qubit complexity as the minimal
number of qubits needed to solve the problem by a hybrid algorithm 
which performs classical and quantum operations with possible 
many measurements. This minimum is, however, zero since we could simulate
all quantum operations on a classical computer with no qubits but
at exponential cost of classical operations. 

We choose to study the qubit complexity of quantum algorithms with one
measurement to eliminate such a possibility.
\ \qed
\vskip 1pc

\vskip .5pc
\noindent {\bf Example\,: Multivariate Integration (continued)}\
\newline
For the class $F=F_{d,r}$  with $r\ge1$,
the minimal error $\eqs(A_{n,k})$ of quantum algorithms $A_{n,k}$ is
of order $n^{-1-r/d}$  and is achieved by an algorithm that
uses of order $\log\,\e^{-1}$ qubits. 
This result follows from reduction of the integration problem to
real and then to Boolean summation, and from the fact 
that the Boolean summation algorithm of \cite{brassard} using $n$ bit
queries with seven repetitions has the error for worst $f$ and average
$j$ also proportional to $n^{-1}$ as proved in \cite{hkw}. 
This implies that 
\begin{eqnarray}
\cques(\e,S)\,=\,\Theta\left(\e^{-1/(1+r/d)}\right) \quad\mbox{and}
\quad\cqubs(\e,S)\,=\,O\left(\log\,\e^{-1}\right).\label{qubrge1}
\end{eqnarray}
\qed

\subsection{Lower Bounds on Qubit Complexity}

We now prove lower bounds on the qubit complexity in the
standard quantum setting in terms of the non-adaptive information
complexity in the worst case setting as well as in terms of the (Kolmogorov) $\e$-entropy of the set $S(F)$.
Based on these bounds, we conclude that some continuous problems $S$ {\it cannot}
be solved in the standard quantum setting.
  
\begin{thm}\label{thm1}
$$
\cqubs(\e,S)\,\ge\,\log\,\cw(2\e,S).
$$
\end{thm}
\vskip 1pc
\noindent {\it Proof.\ }
\newline
Take an arbitrary quantum algorithm $A_{n,k}$ such
that $\eqs(A_{n,k})\le\e$ with the minimal number of qubits $k=\cqubs(\e,S)$.  
We have 
$$
\e^2\,\ge\,\eqs(A_{n,k})^2\,=\,\sup_{f\in F}\,
\sum_{j=0}^{2^k-1}p_{f,j}\|S(f)-A_{n,k}(f,j)\|^2.
$$

The final state as well as probabilities $p_{f,j}$ of the quantum
algorithm $A_{n,k}$ are based on the non-adaptive information
$$
N(f)\,=\,[f(t_1),f(t_2),\dots,f(t_{2^k})]
$$
for some sample points $t_j\in D_d$. Therefore we can write 
$$
A_{n,k}(f,j)\,=\,\Phi(f(t_1),f(t_2),\dots,f(t_{2^k});j)\qquad\forall\,f\in
F,\ \forall\,j\in\{0,1,\dots,2^k-1\},
$$
for some mapping $\Phi:\reals^{2^k}\times \{0,1,\dots,2^k-1\}\to G$.

For an arbitrary $f\in F$, take two functions $f_1$ and $f_2$ such
that $N(f_1)=N(f_2)=N(f)$. The final
state as well as all probabilities $p_{f,j}$ will be the same for $f_1$
and $f_2$, and $a_j=A_{n,k}(f_1,j)=A_{n,k}(f_2,j)$ for all $j$. 
Hence, for any $f\in F$, we have 
$$
2\e^2\,\ge\,\sum_{j=0}^{2^k-1}p_{f,j}\left(\|S(f_1)-a_j\|^2+\|S(f_2)-a_j\|^2\right).
$$
Observe that
$$
\|S(f_1)-S(f_2)\|^2\,\le\,\left(\|S(f_1)-a_j\|+\|S(f_2)-a_j\|\right)^2\,\le\,
2\left(\|S(f_1)-a_j\|^2+\|S(f_2)-a_j\|^2\right).
$$
Multiplying both sides by $p_{f,j}$ and summing up with respect to $j$, we conclude
$$
\|S(f_1)-S(f_2)\|^2\,\le 4\e^2.
$$
Taking the supremum with respect to $f\in F$ and $f_1,f_2$ from $F$
with $N(f_1)=N(f_2)$ we have
$$
\,\sup_{f\in F}\ \sup_{f_1,f_2\in
  F,\,N(f_1)=N(f_2)=N(f)}\ \|S(f_1)-S(f_2)\|\,\le\,2\e.
$$
The left-hand side of the last inequality is equal to the
diameter of information~$N$, see \cite{traub} p. 45, which in turn is
bounded from below by the radius of information, denoted by 
$\mbox{rad}(N)$. Hence, $\mbox{rad}(N)\le 2\e$ which can hold only if
the cardinality  of $N$ is at least equal to $\cw(2\e,S)$, see
\cite{traub} p. 54. Thus, $2^k\ge \cw(2\e,S)$, as claimed.

We add in passing that for many cases we have $d(N)=2{\rm
rad}(N)$. This holds, in particular, if $G=\reals$. Then
${\rm rad}(N)\le \e$ and the extra factor $2$ can be omitted.  
\qed 
\vskip 1pc 
Theorem \ref{thm1} states that the number of qubits needed to solve
$S$ in the standard quantum setting is related to the 
non-adaptive information complexity in the worst case setting.  
For most continuous problems $S$, the 
non-adaptive information complexity goes to infinity as $\e$ approaches
zero. Then Theorem \ref{thm1} says that the number of qubits also goes
to infinity although much more slower due to the presence of the
logarithm. We illustrate this point by continuing our example.
\vskip .5pc
\noindent {\bf Example\,: Multivariate Integration (continued)}\
\newline
Consider $F=F_{d,r}$ with $r\ge1$. We know that
$$
\cw(2\e,S)\,=\,\Theta\left(\e^{-d/r}\right).
$$
Then Theorem \ref{thm1} supplies a lower bound
on the qubit complexity,  
$$
\cqubs(\e,\intt_{d,r})\,\ge \,\tfrac{d}r\,\log\,\e^{-1} \, +\, \Omega(1)
$$ 
with the term $\Omega(1)$ independent of $\e$ but dependent on $d$ and
$r$.

As we already discussed, $\tfrac{d}{r}$ was the exponent of the worst case
complexity of the integration problem $\intt_{d,r}$ and caused the
curse of dimensionality.  Its role for the qubit complexity
is mitigated since it effects the lower bound on the qubit complexity only
linearly. 

Due to (\ref{qubrge1}) the lower bound on the qubit complexity is
sharp with respect to $\e$, and we have 
$$
\cqubs(\e,\intt_{d,r})\,=\,\Theta\left(\log\,\e^{-1}\right).
$$
The dependence on $\e$ is very weak although for $\e$
tending to zero, the qubit complexity slowly goes to infinity.  
\qed
\vskip 1pc
\noindent{\bf Remark 3:}
Theorem 3.1 was formally proved for the class 
of quantum algorithms with one measurement. We now show that
a similar result  holds for the much more larger class 
of hybrid algorithms which use non-adaptive or adaptive function
values on a classical computer and many
measurements on a quantum computer. 
More precisely, consider the following class
of hybrid algorithms:
\begin{itemize}
\item For $i=1,2,\dots,p$ do
\begin{itemize}
\item Use a classical algorithm with $m_i$ non-adaptive or adaptive
      function values to get an initial state $|\psi_{0,i}\rangle$,
\item Use a quantum algorithm $A_{n_i,k_i}$ with one measurement
      starting with the initial state $|\psi_{0,i}\rangle$ and
      with $n_i$ bit queries and $k_i$ qubits, and let  
$$
A_{n_i,k_i}(f,j)\,=\,\phi_i(j)
$$
with a function $\phi_i$ which can now be dependent on $\ell_i$
non-adaptive or adaptive function values. 
\end{itemize}
\end{itemize}
Observe that the total number of function values used by the hybrid
algorithm is at most 
$$
N\,:=\,\sum_{i=1}^p\left(m_i+2^{k_i}+\ell_i\right),
$$
and up to $\sum_{i=1}^p(m_i+\ell_i)$ of them 
can be computed adaptively. The hybrid
algorithm uses $k$ qubits, where
$$
k\,=\,\max_{i=1,2,\dots,p}k_i.
$$
Assume that the error of the hybrid algorithm is $\e$. Then 
as in the proof of Theorem 3.1 we conclude
$$
N\,\ge \cw(2\e,S),
$$
where now $\cw(2\e,S)$ stands for the worst case (adaptive) information
complexity defined as in (\ref{non-adaption}) with the exception that now
$A_n^{\rm nad}$ is replaced by an arbitrary algorithm $A_n$ using
at most $n$ adaptive function values. 

Hence, as long as there are two positive numbers $a_1$ and $a_2$ such
that
$$
N\,\le\,a_1\,2^{\,a_2\,k} 
$$
then
$$
k\,\ge\,a_2^{-1}\log\,\cw(2\e,S)\ -\ \log\,a_1.
$$
Hence, even for hybrid algorithms, the logarithm of the worst case
(adaptive) information complexity is essential and 
tells us how many qubits are needed. \qed 
\vskip 1pc
We now consider the case when $\cw(\e,S)=\infty$, i.e., when we
cannot solve the problem in the worst case setting. 
Then the qubit complexity is
also infinite. We summarize this fact in the following corollary.

\begin{cor}\label{cor1}
If the non-adaptive information complexity of $S$ in the worst case setting
is infinity then $S$ cannot be solved in the standard quantum setting.
\end{cor}
We illustrate Corollary \ref{cor1} by multivariate integration
for $r=0$.
\vskip .5pc
\noindent {\bf Example\,: Multivariate Integration (continued)}\
\newline
Assume now that $r=0$. Hence, $F=F_{d,0}$ is the unit ball of continuous
functions with the norm $\|f||_0=\max_{x\in[0,1]^d}|f(t)|$ bounded by one.
It is known, and easy to see, that for any algorithm $A_n$ we have
$$
\ewor(A_n)\,\ge\,1\qquad\forall\,n.
$$
Indeed, as already explained in the introduction, 
it is enough to take two continuous functions from $F$ vanishing at the
sample points $t_j$ used by the algorithm $A_n$, $j=1,2,\dots,n$,
such that the integral of the first function is almost $1$, and the integral of
the other function is almost $-1$. Since these functions
are indistinguishable for the algorithm $A_n$, the best we can do is
to approximate their integrals by zero with error arbitrarily close 
to one. Hence, the worst case error of any algorithm is at least one,
as claimed. This implies that 
$$
\cw(\e,S)\,=\,\infty\qquad\forall\,\e<1.
$$
Theorem \ref{thm1} says that multivariate integration for $r=0$ is
{\it unsolvable} in the standard quantum setting.

As already mentioned, this problem is, however, easy in the randomized
setting. The randomized error of the Monte Carlo is bounded
by $n^{-1/2}$ which is optimal due to lower bounds of Bakhvalov and Novak.
Therefore the non-adaptive randomized information complexity
as well as the total randomized complexity are both of order $\e^{-1/2}$.

So why does the standard quantum setting fail for the problem which is
relatively easy in the randomized setting? As we shall see in the next
section, the reason is that we use {\it deterministic} queries
in the standard quantum setting. This bad property will disappear if
we allow the use of {\it randomized} queries also
in the quantum setting. \ \qed
\vskip 1pc
Before we proceed to the quantum setting with randomized queries, we
briefly present another lower bound on the qubit complexity in the
standard quantum setting. This bound relates the qubit complexity
to the (Kolmogorov) $\e$-entropy of the set $S(F)$. We first recall the notion of 
$\e$-entropy in normed spaces, see e.g., \cite{Lorentz}. 
Let $B$ be a subset of $G$. We want to
cover the subset $B$ by the minimal number of subsets of $G$ whose
diameters do not exceed~$2\e$. That is, let
$$
n(\e,B)\,=\,\min\left \{\,n\,:\ \exists\,B_j\subset G\ \mbox{such that\ } 
{\rm diam}(B_j)\le2\e,\ B\subset\cup_{j=1}^nB_j\ \right\},
$$
where ${\rm diam}(B_j)=\sup_{b_1,b_2\in B_j}\|b_1-b_2\|$. Then
the $\e$-entropy of $B$ is
$$
\mbox{Ent}(\e,B)\,=\,\log\,n(\e,B).
$$
It is easy to prove the following theorem.  
\begin{thm}\label{thm2}
$$
\cqubs(\e,S)\,\ge\,{\rm Ent}(\e,S(F)).
$$
\end{thm}
\vskip 1pc
\noindent {\it Proof.\ } 
\newline
The proof relies on the fact that in the
standard quantum setting any quantum algorithm 
which uses $k=\cqubs(\e,S)$ qubits produces
at most $2^k$ different elements $A_{n,k}(f,j)=\phi(j)$ from $G$
for $j=0,1,2,\dots,2^k-1$ with $\phi$ independent of $f$. 

We take an arbitrary 
quantum algorithm $A_{n,k}$ with error
$\eqs(A_{n,k})\le \e$. For any $f\in F$, we have
$$
\min_{j=0,1,\dots,2^k-1}\|S(f)-\phi(j)\|\,\le\,
\left(\sum_{j=0}^{2^k-1}p_{f,j}\|S(f)-\phi(j)\|^2\right)^{1/2}\,\le\,\e.
$$
Let $B(\phi(j),\e)=\{g\in G\,:\, \|g-\phi(j)\|\le\e\,\}$
be the ball in $G$ of center $\phi(j)$ and radius~$\e$.
Obviously, ${\rm diam}(B(\phi(j),\e))\le 2\e$. 
Then $S(f)\in\cup_{j=0}^{2^k-1}B(\phi(j),\e)$
and therefore $S(F)\subset
\cup_{j=0}^{2^k-1}B(\phi(j),\e)$.
This means that $2^k\ge n(\e,S(F))$,
and $k\ge \log\,n(\e,S(F))$, as claimed. \qed
\vskip 1pc
The essence of Theorem \ref{thm2} is that $S(F)$ must have a finite
$\e$-entropy in order to have $S$ solvable in the standard quantum
setting. In particular, this means that the closure of $S(F)$ must be compact.
Otherwise, the $\e$-entropy is infinite and the finite number of
qubits is not enough to solve $S$. We summarize this in the following
corollary.
\begin{cor}\label{cor2}
If the closure of $S(F)$ is not compact then
 $S$ is not solvable in the standard quantum setting. 
In particular, if $S(F)$ is unbounded then
 $S$ is not solvable in the standard quantum setting. 
\end{cor}
The unboundedness of $S(F)$ can happen even for problems with 
relatively small worst case complexity as shown in the following example.
This example also shows that lower bounds based on the $\e$-entropy of
$S(F)$  
presented in Theorem \ref{thm2} may be
quite different than lower bounds based on the non-adaptive information
complexity in the worst case setting presented in Theorem \ref{thm1}. 
 
\vskip .5pc
\noindent {\bf Example: Unbounded $S(F)$}\
\newline
Consider the univariate integration problem for Lipschitz functions,
i.e.,
$$
F\,=\,\{\,f:[0,1]\to\reals|\ \ |f(x)-f(y)|\le |x-y|\ \
\forall\ x,y\in[0,1]\ \},
$$
and $S(f)=\int_0^1f(t)\,dt$ with $G=\reals$. 

Since all constant functions belong to $F$, we have $S(F)=\reals$ and
therefore $\mbox{Ent}(\e,\reals)=\infty$. It is well known that 
the worst case complexity is roughly $1/(4\e)$, see e.g., \cite{TW98},
and the linear algorithm
$$
A_n(f)\,=\,\frac1{2n}f\left(\frac1{2n}\right)\,+\,\frac1n\sum_{j=2}^{n-1}
f\left(\frac{2j-1}{2n}\right)\,+\,
\frac1{2n}f\left(\frac{2n-1}{2n}\right)
$$
with $n=\lceil \e^{-1}/4\rceil$ minimizes the worst case error
among all algorithms using $n$ function values, and has error at most
$\e$. 
Observe that for constant functions, $f(t)\equiv c$, we have
$A_n(f)=c$ which may be arbitrary large. 

In the worst case
setting with the real number model, the sizes of numbers
do not matter and do not affect the cost analysis. In the standard
quantum setting, the situation is different since we can only work on
unit vectors and the scaling of numbers {\em does} matter. That is why 
we cannot solve unscaled problems in the standard quantum setting.

In many cases, we may rescale the problem by changing $F$ to
a set $\tilde F$ such that $S(\tilde F)$ is bounded and its closure 
is compact. This idea works for our example as follows.
For $f\in F$, define $g(x)=f(x)-f(0)$ for $x\in [0,1]$. Then 
$g(0)=0$ and $|g(x)|\le x\le1$. Hence, $-x\le g(x)\le x$ and
therefore $S(g)\in[-\tfrac12,\tfrac12]$, and both bounds are sharp. 
Define
$$
\widetilde F\,=\,\{g:[0,1]\to\reals|\ g(0)=0,\ g\in F\}.
$$
Then $f\in F$ iff $g\in\widetilde F$ for $g(x)=f(x)-f(0)$, and
$S(f)=S(g)-f(0)$. 

We now have $S(\widetilde F)= [-\tfrac12,\tfrac12]$ and therefore 
$$
{\rm Ent}(\e,S(\widetilde F))\,=\,\log\,\e^{-1}+O(1).
$$
Hence, the number of qubits for approximations of $S(g)$ 
is now bounded by roughly $\log\,\e^{-1}$. In fact it is sharp,
since $S(g)$ can be approximated to within $\e$ 
in the standard quantum setting by $A_{n,k}(q)$ using roughly $n=\e^{-1/2}$ bit
queries  and $k=\log\,\e^{-1}$ qubits as shown in~\cite{N01}.

Finally, we may approximate $S(f)$ for $f\in F$ by running $A_{n,k}(g)$
for $g(x)=f(x)-f(0)$, and computing 
$f(0)+ A_{n,k}(q)$ on a classical computer. Note that the last step on
a classical computer may involve an arbitrarily large number $f(0)$
which is of no relevance as long as we use the real number model of 
computation.
 \ \qed 

\section{Quantum Setting with Randomized Queries}

Modulo measurements, the standard quantum setting for continuous
problems is similar to the worst case setting with non-adaptive
information. All unitary matrices including queries, 
as well as the number of qubits  used by quantum algorithms are 
deterministic. The potential speedup of the standard quantum setting 
for continuous problems over the worst case setting relies
on the fact that quantum algorithms with $k$ qubits may use an exponential
number up to $2^k$ function values with cost proportional to a
small power of $k$. If $2^k$ functions values are not enough to solve
the problem in the worst case setting then the problem remains
unsolvable also in the standard quantum setting. As we indicated
before, this may happen even for problems with small randomized
complexity. Such examples suggest studying more general
quantum settings.

In this section, we describe the quantum setting with 
{\it randomized queries} in which all unitary matrices including queries
may be randomized. 
Modulo measurements, the quantum setting with randomized queries
for continuous
problems will be similar to the randomized setting with non-adaptive
information. We assume that the number of qubits is fixed
and does not depend on randomization. As we already mentioned
in the introduction the extension to randomized qubits
is left for future study.

We generalize (\ref{quantum}) by allowing unitary matrices
$U_j$ as well as the query $Q_f$ to be
randomly chosen similarly as in the randomized classical setting of Section 2.2.
That is, we have random elements $\omega$ distributed accordingly to
some probability measure $\rho$ on $\Omega$ with $\rho(\Omega)=1$.
We stress that $\rho$ does not depend on $f$ and
we will be using the same randomization for all $f$ from $F$. 

First we choose $k$ as the number of qubits, take a random element
$\omega$, choose a unit vector $|\psi_{0,\omega}\rangle$ from ${\cal C}^{2^k}$
as the initial state, and obtain the final $k$ qubit state
\begin{equation}\label{pipi}
|\psi_{f,\omega}\rangle\,=\,U_{n_\omega,\omega}Q_{f,\omega}\,
U_{n_\omega-1,\omega}Q_{f,\omega}\,\cdots\,U_{1,\omega}Q_{f,\omega}\,
U_{0,\omega}|\psi_{0,\omega}\rangle.
\end{equation}
For a fixed $\omega$, we have the same situation as in the standard
quantum setting. That is, matrices $U_{j,\omega}$ are arbitrary
$2^{k}\times 2^{k}$ unitary matrices which are
independent of $f$, and the query
$Q_{f,\omega}$, which is also a $2^{k}\times 2^{k}$
unitary matrix, depends on at most $2^{k}$ sample points
which are independent of $f$ and depend only on $\omega$,
$$
Q_{f,\omega}\,=\,Q_{f(t_{1,\omega}),f(t_{2,\omega}),\dots,f(t_{2^k,\omega})}.
$$
Hence, $Q_{f,\omega}$ is a randomized query depending on at most
$2^{k}$ function values at randomized sample points.
In full analogy with the standard quantum setting, 
the measure $\rho$ and sample points $t_{j,\omega}$ are
the same for all $f$ from $F$. That is, we use {\it non-adaptive}
randomized information with fixed cardinality at most $2^k$.  Let
$$
n\,=\,\int_{\Omega}n_\omega\,\rho(d\omega)
$$
be the average number of queries used to obtain the final states. 

We then perform a measurement which for a fixed $\omega$ is the same
as for the standard quantum setting. That is, we obtain an index 
$j\in\{0,1,\dots,2^{k}-1\}$ with probability
$p_{f,j,\omega}$ depending on the final state
$|\psi_{f,\omega}\rangle$,
where $\sum_{j=0}^{2^k-1}p_{f,j,\omega}=1$ for all $\omega\in\Omega$. 
As before, the dependence on $f$ is only through function values,
$$
p_{f,j,\omega}\,=\,p_{f(t_{1,\omega}),f(t_{2,\omega}),\dots,
f(t_{2^{k}},\omega),j,\omega}.
$$ 
Knowing the index $j$, we compute on a classical computer
$$
A_{n,k}(f,\omega,j)\,=\,\phi_\omega(j)
$$
for some mapping $\phi_\omega:\{0,1,\dots,2^{k}-1\}\,\to\,G$. 
The algorithm $A_{n,k}$ is called a {\em quantum
algorithm using randomized queries}, or just a quantum algorithm
if it is clear from the context that we are using randomized queries.

Analogously to the standard quantum setting, we consider 
the error of a quantum
algorithm by taking the average performance with respect to both
$j$ and $\omega$, see also the appendix where
the probabilistic error is discussed.  
That is, the {\em error} of an algorithm in the
{\em quantum setting with randomized queries} 
$A_{n,k}$ is  defined by
$$
\eqr(A_{n,k})\,=\,\sup_{f\in F}\ \left(\int_{\Omega}
\sum_{j=0}^{2^{k}-1}p_{f,j,\omega}\|S(f)-A_{n,k}(f,\omega,j)\|^2\rho(d\omega)\right)^{1/2}. 
$$
Observe that if we choose all matrices $U_{j,\omega}$ and
$Q_{f,\omega}$ as well as $n_\omega$ independently of $\omega$, then
this definition coincides with the error  
in the standard quantum setting. 

This definition of the error leads to the {\it randomized query
complexity} defined by 
$$
\cquer(\e,S)\,=\,\min\left\{\,n\,:\ \exists\,A_{n,k}\ \mbox{such that\
  } \eqr(A_{n,k})\,\le\,\e\ \right\},
$$
and to the {\it randomized qubit complexity} 
defined by 
$$
\cqubr(\e,S)\,=\,\min\left\{\,k\,:\ \exists\,A_{n,k}\ \mbox{such that\
  } \eqr(A_{n,k})\,\le\,\e\ \right\}.
$$

As in the standard quantum setting, we may have a tradeoff 
between the minimal number of
queries and qubits. Therefore it would be also reasonable to study 
the minimization of the product or a weighted sum of queries and qubits
in the quantum setting with randomized queries, however, we do not
pursue the issue in this paper.

It is natural to ask what kind of results can be now achieved 
and how much we can improve the results
from the standard quantum setting. We will study these questions in
the next sections.
\subsection{Lower Bounds on Randomized Qubit Complexity}

For the standard quantum setting, we proved lower bounds on qubit
complexity in terms of the worst case setting on a classical computer.
We now show that 
lower bounds on the randomized qubit complexity 
can be analogously derived in terms of the randomized setting on a classical computer. 
\begin{thm}\label{thm1ran}
$$
\cqubr(\e,S)\,\ge\,\log\,{\rm comp}^{{\rm inf-ran}}(\e,S).
$$
\end{thm}
\vskip 1pc
\noindent {\it Proof.\ } 
\newline
We note that any quantum algorithm $A_{n,k}$ can be regarded as a
randomized algorithm whose cardinality is at most $2^k$. 
Indeed, for $f\in F$, let $\bar{\rho}_f$ be a probability measure
defined on $B\times J$, where $B$ is a
measurable subset of $\Omega$ and $J$ is an arbitrary subset of
$\{0,1,\dots,2^k-1\}$ given by 
$$
\bar{\rho}_f(B\times J)\,=\,\int_{\Omega}1_{B}(\omega) \sum_{j\in
  J}p_{f,j,\omega}\, \rho(d\omega)
$$
with the characteristic function $1_B(\omega)=1$ for $\omega\in B$ and
$1_B(\omega)=0$ otherwise. Then
\begin{eqnarray*}
\eqr(A_{n,k})^2\,&=&\,
\sup_{f\in F}\ \int_{\Omega}
\sum_{j=0}^{2^{k}-1}p_{f,j,\omega}\|S(f)-A_{n,k}(f,\omega,j)\|^2\rho(d\omega)\\
&=&\,
\sup_{f\in F}\ \int_{\Omega\times\{0,1,\dots,2^k-1\}}
\|S(f)-A_{n,k}(f,\omega,j)\|^2\bar{\rho}_f(d(\omega,j))\,=\,
e^{{\rm ran}}(A_{n,k})^2.
\end{eqnarray*}
Hence, $A_{n,k}$ can be regarded as a randomized algorithm 
in the randomized classical setting whose cardinality is at most
$2^k$. Since the sample points $t_{j,\omega}$ are independent of $f$
and dependent only on $\omega$, the algorithm $A_{n,k}$ uses
non-adaptive information, and $A_{n,k}$ applied to $f$ uses randomization with 
the measure $\bar{\rho}_f$. 
If $\eqr(A_{n,k})\le \e$ 
then $e^{{\rm ran}}(A_{n,k})\le \e$ which may happen only if  
the cardinality of $A_{n,k}$ is at least 
$\crr(\e,S)$. This means that $2^k\ge\crr(\e,S)$, as claimed.\ \qed
\vskip 1pc
Note that the lower bounds in Theorem \ref{thm1ran} 
are not larger than the lower bounds in Theorem \ref{thm1}
since ${\rm comp}^{{\rm inf-ran}}(\e,S)\le
{\rm comp}^{{\rm inf-wor}}(\e,S)$. Furthermore for some
problems the non-adaptive information complexity in the worst case
setting may be infinite whereas its randomized counterpart is
relatively small. 

We already mentioned that the integration problem for the class
$F=F_{d,0}$ is unsolvable in the standard quantum setting and
solvable in the randomized classical setting. We now provide a proof and
find the randomized qubit complexity. The randomized query complexity
for $F_{d,0}$ as well as the integration problem for $F_{d,r}$ for an arbitrary
integer $r\ge0$ will be studied later. 
\vskip .5pc
\noindent {\bf Example\,: Multivariate Integration for $r=0$ (continued)}\
\newline
We show that the randomized qubit complexity for $F=F_{d,0}$ is
$$
\cqubr(\e,\intt_{d,0})\,=\,\Theta\left(\log\,\e^{-1}\right).
$$
and is achieved by a quantum algorithm which uses of order
$\e^{-1}$ queries.

We know that ${\rm comp}^{{\rm
    inf-ran}}(\e,\intt_{d,0})=\Theta(\e^{-2})$.
{}From Theorem \ref{thm1ran} we conclude that the randomized qubit complexity must
be at least of order $\log\,\e^{-1}$. 

We now provide an upper bound on the randomized qubit complexity. Take the Monte Carlo
algorithm with $m=\lceil 4\e^{-2}\rceil $,
$$
{\rm MC}_m(f,\omega)\,=\,\frac1m\sum_{j=1}^mf(\omega_j)
$$
with $\omega=[\omega_1,\omega_2,\dots,\omega_m]$ and independent
uniformly distributed $\omega_j$ from $[0,1]^d$. 
Then $e^{{\rm ran}}({\rm MC}_m)\le\e/2$. 

We now apply the Boolean summation algorithm $A^*_{n,k}$ of
\cite{brassard} with seven repetitions for
real functions for which $|f(x)|\le 1$, see also Section 5.
The algorithm $A^*_{n,k}$ approximates ${\rm MC}_m(f)$.
It is known, see \cite{hkw},
that the randomized error of this algorithm is bounded by $C/n$,
where $C$ is a number independent on $f,n$ and $m$.
Furthermore, the algorithm uses 
$k=\Theta(\log\,m)$ qubits. We set $n=\lceil 2\,C/\e\rceil$
and obtain
$$
\sum_{j=0}^{2^k-1}p_{f,j,\omega}\left[{\rm
    MC}_m(f,\omega)-A^*_{n,k}(f,\omega,j)\right]^2\,
\le\,\frac{\e^2}4\qquad\forall\,f\in F.
$$
Therefore for any $f\in F$ we have 
\begin{eqnarray*}
&&\sum_{j=0}^{2^k-1}p_{f,j,\omega}\bigg[\intt_{d,0}(f)-A^*_{n,k}
(f,\omega,j)\bigg]^2\,=\,\\
&&\sum_{j=0}^{2^k-1}p_{f,j,\omega}\bigg[\intt_{d,0}(f)-{\rm
    MC}_m(f,\omega)+
{\rm  MC}_m(f,\omega) -A^*_{n,k}(f,\omega,j)\bigg]^2\,\le\,\\
&&2\,\sum_{j=0}^{2^k-1}p_{f,j,\omega}\bigg[\left(\intt_{d,0}(f)-{\rm
    MC}_m(f,\omega)\right)^2\,+\,\left(
{\rm  MC}_m(f,\omega) -A^*_{n,k}(f,\omega,j)\right)^2\bigg]\,\le\,\\
&&\,2\bigg[\intt_{d,0}(f)-{\rm
    MC}_m(f,\omega)\bigg]^2\,+\,\frac{\e^2}2.
\end{eqnarray*}
Taking the integral over $\Omega$ we conclude that
$$
e^{{\rm ran}}(A^*_{n,k})\,\le \,\sqrt{2\frac{\e^2}4+\frac{\e^2}2}\,=\,\e.
$$
Hence, we can solve the integration problem for $r=0$ using
of order $\log\,\e^{-1}$ qubits and $\e^{-1}$ randomized
queries, as claimed. \ \qed
\vskip 1pc
Theorem \ref{thm1ran} states that the number of qubits in the
quantum setting with randomized queries 
depends on the non-adaptive information 
complexity ${\rm comp}^{{\rm inf-ran}}(\e,S)$. Typically, 
${\rm comp}^{{\rm inf-ran}}(\e,S)$  goes to infinity with $\e$
tending to zero, so the number of qubits has to go to infinity
as well although much slower due to the presence of the logarithm in
the bound of Theorem \ref{thm1ran}. However, if
${\rm comp}^{{\rm inf-ran}}(\e,S)=\infty$ then the randomized qubit complexity 
is infinity and the problem cannot be solved. We summarize this fact
in the following corollary.
\begin{cor}\label{corran}
If the non-adaptive information complexity of $S$ in the randomized setting
is infinity then $S$ cannot be solved in the quantum setting with
randomized queries.
\end{cor}
\vskip 1pc
We illustrate Corollary \ref{corran} by a problem with infinite
non-adaptive information complexity in the quantum setting
with randomized queries.
\vskip 1pc
\noindent {\bf Example\,: Multivariate Approximation}\ 
\newline
Consider the same class $F=F_{d,r}$, $r\ge0$,  as for the multivariate
integration problem. Let $G=C([0,1]^d)$ and $S={\rm
  APP}_{d,r}:C^r([0,1]^d)\to G$ be defined by
$$
{\rm APP}_{d,r}(f)\,=\,f.
$$
It is known, see \cite{traub} p. 425, that randomization does not help
for this problem and that for small $\e$ we have
$$
{\rm comp}^{{\rm inf-ran}}(\e,{\rm APP}_{d,0})\,=\,
{\rm comp}^{{\rm inf-wor}}(\e,{\rm APP}_{d,0})\,=\,\infty,
$$
and for $r\ge1$,
$$
{\rm comp}^{{\rm inf-ran}}(\e,{\rm APP}_{d,r})\,=\,\Theta\left(
{\rm comp}^{{\rm inf-wor}}(\e,{\rm APP}_{d,r})\right)\,=\,\Theta\left(\e^{-d/r}\right).
$$
Hence, for $r=0$ we conclude that the approximation problem cannot be
solved in the quantum setting with randomized queries. \ \qed

\section{Boolean and Real Summation}

Solution of the real summation problem is a basic module used in the 
solution of many continuous problems. 
The real summation problem is very much related 
to the Boolean summation problem which has been 
thoroughly studied in the standard quantum setting, see
\cite{brassard,heinrich,hkw,kw,nayak}. In this section we study the
Boolean and real summation problems in the quantum setting
with randomized queries. 

\subsection{Boolean Summation}

For $N$ a (large) power of two, consider the class of Boolean
functions
$$
F\,=\,F_N\,=\,\{f\,|\ \ f:\{0,1,\dots,N-1\}\,\to\,\{0,1\}\ \},
$$
and take $G=\reals$. The Boolean summation problem is defined as 
$S={\cal B}_N:F_N\to G$ given by
$$
{\cal B}_N(f)\,=\,\frac1n\sum_{j=0}^{N-1}f(j).
$$
We want to approximate ${\cal B}_N$ to within $\e$. Without loss of
generality we may assume that $\e\ge1/(2N)$. Indeed, if we know
$A_{n,k}(f)$ such that $|{\cal B}_N(f)-A_{n,k}(f)|\le\e<1/(2N)$ then
we can recover ${\cal B}_N(f)$ exactly since
$$
{\cal B}_N(f)\,=\,\frac{\lceil N\,A_{n,k}(f)+\tfrac12\rceil\,-\,1}{N}.
$$
This follows from the fact that we know a priori that ${\cal
  B}_N(f)=k/N$ for some integer $k\in[0,N]$ with $k$ being the total
number of the true assignments of the Boolean function $f$. Then
$NA_{n,k}(f)+\tfrac12=k+x$ with
$$
x\,=\,\tfrac12+N\left(A_{n,k}(f)-{\cal
    B}_N\right)\,\in\,\left[\tfrac12-N\e,\tfrac12+N\e\right]\,
\subset\,(0,1).
$$
Hence, $\lceil k+x\rceil=k+1$, as claimed.

We first consider the Boolean summation problem in the standard
quantum setting.
The Boolean summation algorithm $A^*_{n,k}$ of \cite{brassard} 
with seven repetitions solves the Boolean summation such that
$$
e^{{\rm qua-std}}(A_{n,k}) \,\le\,\e,
$$
using $n=\Theta(\e^{-1})$ bit queries and $k=\Theta(\log\,N)$ qubits. 
The query bound is order-optimal, see \cite{hkw,nayak}. 
The qubit bound is also order-optimal since it
is known, see e.g., \cite{N01},  that in the worst case setting
$$
{\rm comp}^{{\rm inf-wor}}(\e,S)\,=\,\lceil
N(1-2\e)\rceil\qquad\forall\,
\e\in\left[0,\tfrac12\right]
$$
which is essentially $N$ for small $\e$. From Theorem \ref{thm1} we
conclude that the qubit complexity is
roughly at least $\log\,N$ for small $\e$. 
We summarize these results in the following theorem.
\begin{thm}\label{thmboolstd}
The complexities of the Boolean summation problem in the standard
quantum setting satisfy 
$$
{\rm comp}^{{\rm que-std}}({\cal B}_N)\,=\,\Theta\left(\e^{-1}\right)\qquad
{\rm comp}^{{\rm qub-std}}({\cal B}_N)\,=\,\Theta\left(\log\,N\right).
$$
Furthermore, these bounds are both attained by the Boolean
summation algorithm with seven repetitions.
\end{thm}
\vskip 1pc
We now consider the Boolean summation problem in the 
quantum setting with randomized queries. 
We prove the following theorem. 

\begin{thm}\label{thmboolran}
The complexities of the Boolean summation problem in the 
quantum setting with randomized queries satisfy
$$
{\rm comp}^{{\rm que-ran}}({\cal B}_N)\,=\,\Theta\left(\e^{-1}\right)\qquad
{\rm comp}^{{\rm qub-ran}}({\cal B}_N)\,=\,\Theta\left(\log\,\e^{-1}\right).
$$
Furthermore, these bounds are both attained by the Boolean
summation algorithm with seven repetitions applied to  
$$
\frac1m\sum_{j=1}^mf(\omega_j)
$$ 
with $m=\Theta(\e^{-2})$ and with independent uniformly
distributed $\omega_j$ from $\{0,1,\dots,N-1\}$.
\end{thm}
\vskip 1pc
\noindent {\it Proof.\ } 
\newline
We first consider lower bounds and start with the randomized query complexity.
We use the known proof technique of using the average case error as a
lower estimate of the randomized error. More precisely, take an arbitrary quantum
algorithm $A_{n,k}$ that uses $n$ randomized bit queries and $k$
qubits, 
and consider its randomized error
$$
\eqr(A_{n,k})^2\,=\,\sup_{f\in F_N}\ \int_{\Omega}
\sum_{j=0}^{2^{k}-1}p_{f,j,\omega}|{\cal B}_N(f)-A_{n,k}(f,\omega,j)|^2\rho(d\omega). 
$$
We now replace the supremum over $f$ by an average over $f$. That is,
we assume that a Boolean function $f$ from $F_N$ occurs with
probability $p_f$ with non-negative $p_f$ such that $\sum_{f\in
  F_N}p_f=1$. Observe that this is a well defined measure since $F_N$
consists of finitely many Boolean functions, in fact, we have $2^{N}$
functions in $F_N$.
Then
\begin{eqnarray}
\eqr(A_{n,k})^2\,
&\ge&\,\sum_{f\in F_N}p_f \int_{\Omega}
\sum_{j=0}^{2^{k}-1}p_{f,j,\omega}|{\cal
  B}_N(f)-A_{n,k}(f,\omega,j)|^2\rho(d\omega)\nonumber \\
&=&\, \int_{\Omega}\left(
\sum_{f\in F_N}p_f\sum_{j=0}^{2^k-1}p_{f,j,\omega}|
{\cal B}_N(f)-A_{n,k}(f,\omega,j)|^2\right)\rho(d\omega).\label{777}
\end{eqnarray}
The following result proved by Papageorgiou in \cite{Papa} will be
needed for our consideration. Take the uniform distribution for
${\cal B}_N(f)$, i.e., $p_f={N\choose j}/2^N$ for all $f$ with ${\cal
  B}_N(f)=j/N$. Then there are two positive numbers $c_1$ and $c_2$
with the following properties. For any algorithm $A_{n,k}$ in the
standard quantum setting with $n$ bit queries, 
such that $n\le c_1N$, and $k$ qubits, let
$p_{f,j}$ denote the probability of obtaining the index $j$ through measurement. Let 
$$
\mu_f(J)\,=\,\sum_{j\in J}p_{f,j}
$$
denote the probability of a subset $J$ of $\{0,1,\dots,2^k-1\}$. Then
it is proved in \cite{Papa} that 
$$
\sum_{f\in F_N}p_f\,\mu_f\left(\{j:\ |{\cal B}_N(f)-A_{n,k}(f,j)|\,\ge\,c_2/n\,\}\right)
\,\ge\,0.25.
$$
{}From Chebyshev's inequality we conclude that
$$
\sum_{f\in F_N}p_f\,\sum_{j=0}^{2^k-1}|{\cal
  B}_N(f)-A_{n,k}(f,j)|^2\,\ge\,\frac14\left(\frac{c_2}n\right)^2.
$$
We apply the last inequality for an arbitrary algorithm
$A_{n,k}$ from the quantum setting with randomized queries and
with a fixed $\omega$.
Here, we use the fact that the algorithm $A_{n,k}(\cdot,\omega,\cdot)$
can be regarded as an algorithm from the standard quantum setting.  
We thus have 
$$
\sum_{f\in F_N}p_f\sum_{j=0}^{2^k-1}p_{f,j,\omega}|
{\cal B}_N(f)-A_{n,k}(f,\omega,j)|^2 \ge\,
\frac14\left(\frac{c_2}n\right)^2.
$$
Since the right-hand side is independent of $\omega$, from (\ref{777}) we obtain
$$
\eqr(A_{n,k})\,\ge\,\frac{c_2}{2n}.
$$
Hence,
$\eqr(A_{n,k})\le\e$ implies that $n=\Omega(\e^{-1})$ and
$$
{\rm comp}^{{\rm que-ran}}({\cal B}_N)\,=\,\Omega\left(\e^{-1}\right).
$$

To prove a lower bound on the randomized qubit complexity, we use
Theorem \ref{thm1ran}. In the randomized setting on a classical
computer it is known that the randomized complexity is of order
$\e^{-2}$, see \cite{N01}. Then Theorem \ref{thm1ran} yields 
$$
{\rm comp}^{{\rm qub-ran}}({\cal B}_N)\,=\,\Omega\left(\e^{-1}\right).
$$
We turn to upper bounds. The idea is the same as for
multivariate integration for $r=0$ which was studied before.
That is, we apply the 
Boolean summation algorithm with seven repetitions to the Monte
Carlo algorithm $m^{-1}\sum_{j=1}^{m}f(\omega_j)$
with independently and uniformly distributed $\omega_j$ over
$\{0,1,\dots,N-1\}$. 
We stress that this algorithm uses randomized queries
$Q_{f,\omega}$ and the rest of unitary matrices are deterministic,
i.e., $U_{j,\omega}=U_j$ in (\ref{pipi}).

The same analysis done for 
$r=0$ yields that the randomized error is $\e$. Since this algorithm
uses of order $\e^{-1}$ randomized queries and $\log\,\e^{-1}$ qubits,
we obtain upper bounds which match the lower bounds. 
This completes
the proof. \ \qed.
\vskip 1pc
It is interesting to compare the complexities of the Boolean summation
problem in the quantum settings with deterministic and
randomized queries. As we see, the query
complexities are roughly the same in both settings. The qubit complexities,
however,  are quite different. For deterministic queries,
the number of qubits depends on the common domain of Boolean
functions, and we need 
roughly $\log\,N$ qubits which can be arbitrary large for large $N$.
For randomized queries, the number of qubits does
not depend on the common domain of Boolean functions. It depends on
the error parameter through $\log\,\e^{-1}$. As we shall see in the
following sections, there is sometimes an exponential difference between 
$\log\,N$ and $\log\,\e^{-1}$. 

\subsection{Real Summation}

We finish this section by a brief note on the real summation
problem.  We now consider $f:\{0,1\dots,N-1\}\to [0,1]$ and want to
approximate
\begin{equation}\label{realsum}
{\rm SUM}_N(f)\,=\,\frac1N\,\sum_{j=0}^{N-1}f(j).
\end{equation}

A known idea is to replace the real number $f(j)$ from $[0,1]$ by its
binary expansion,
$$
f(j)\,=\,\sum_{i=1}^\infty2^{-i}\,f(i,j)\quad\mbox{with} \ f(i,j)\,\in\,\{0,1\},
$$
We define $K=\lceil \log\,\e^{-2}\rceil$ and truncate $f(j)$ to $K$
bits. Let
$$
S_K(f)\,=\, \frac1N\,\sum_{j=0}^{N-1}\sum_{i=1}^K2^{-i}\,f(i,j).
$$  
Clearly, $|{\rm SUM}_N(f)-S_K(f)|\le 2^{-K}\le \e^2$. To obtain a Boolean
function we finally define the set
$$
D\,=\,\left\{(i,j,p)\,: i=1,2,\dots,K,\,j=0,1,\dots,N-1,\,p=1,2,\dots,2^{K-i}\,\right\}
$$
of cardinality $N(2^K-1)$ and a Boolean function $b_f:D\to \{0,1\}$ by
$$
b(i,j,k)\,=\,\delta_{f(i,j),1},
$$
where $\delta_{i,j}$ is the Kronecker delta. Then
$$
\sum_{i=1}^K2^{-i}f(i,j)\,=\,2^{-K}\sum_{i=1}^K2^{K-i}f(i,j)\,=\,2^{-K}\sum_{i=1}^K\sum_{p=1}^{2^{K-i}}b(i,j,p).
$$
Thus
$$
S_K(f)\,=\,{\cal
  B}_{N2^K}(b_f)\,=\,\frac1{N2^K}\,\sum_{j=0}^{N-1}\sum_{i=1}^K\sum_{p=1}^{2^{K-i}}b(i,j,p)
$$
is a Boolean summation problem. We compute ${\cal B}_{N2^K}(b_f)$
with error $\e-\e^2$  by the Boolean summation algorithm with seven
repetitions as explained in Theorem \ref{thmboolran} and obtain
$A_{n,k}(b_f)$ which uses  
of order $\e^{-1}$ randomized queries and
$\log\,\e^{-1}$ qubits. It is easy to check that 
$A_{n,k}(b_f)$ approximates $S(f)$ with the randomized error
at most $\e$. We summarize this in the corollary.
\begin{cor}\label{realran}
The complexities of the real summation problem 
(\ref{realsum}) in the quantum setting with randomized queries satisfy
$$
{\rm comp}^{{\rm que-ran}}({\rm SUM}_N)\,=\,\Theta\left(\e^{-1}\right)\qquad
{\rm comp}^{{\rm qub-ran}}({\rm SUM}_N)\,=\,\Theta\left(\log\,\e^{-1}\right).
$$
Furthermore, these bounds are both attained by the Boolean
summation algorithm with seven repetitions applied to  
$$
\frac1m\sum_{\ell=1}^mf(i_\ell,j_\ell,p_\ell)
$$ 
with $m=\Theta(\e^{-2})$ and with independent uniformly
distributed $(i_\ell,j_\ell,p_\ell)$  over $D$.
\end{cor}

\section{Multivariate Integration}

In this section, we consider the multivariate integration problem 
$\intt_{d,r}$ for the class
$F_{d,r}$ which we used throughout as an illustrative  
example. The purpose of this
section is to study this problem in the quantum settings
with deterministic and randomized queries.

We begin with deterministic queries.
For $r=0$ the problem is unsolvable. For $r\ge1$, 
sharp bounds on the  bit query follow from \cite{N01},
$$
{\rm comp}^{{\rm que-std}}\left(\e,\intt_{d,r}\right)\,=\,
\Theta\left(\e^{-1/(1+r/d)}\right).
$$
This bound is achieved by the Boolean summation algorithm with seven
repetitions 
and uses of order $\log\,\e^{-1}$ qubits, as shown in the previous section.

Observe that the lower bound
on the qubit complexity is of order $\log\,\e^{-1}$ due to 
Theorem~\ref{thm1ran} and the fact that 
the worst case information complexity is of order
$\e^{-d/r}$. Hence, we have
$$
{\rm comp}^{{\rm
    qub-ran}}(\e,\intt_{d,r})\,=\Theta\left(\log\,\e^{-1}\right).
$$

We now turn to randomized queries. The case $r=0$ has
already been covered and we know that we can solve the problem using of
order $\e^{-1}$ randomized bit queries and $\log\,\e^{-1}$
qubits.

For $r\ge1$, we use the same quantum algorithm as in \cite{N01}.
Since the Boolean summation algorithm uses the same order of 
bit queries for the randomized and probabilistic quantum errors,
we obtain the same upper bounds on the number of bit queries and
qubits. 

The lower bound on the number of randomized queries can be derived as
in \cite{N01} and using the results on the Boolean and real 
summation problems of the previous section. This yields that
the randomized bit query complexity  is of order $\e^{-1/(1+r/d)}$.
The lower bound on the number of qubits follows from Theorem
\ref{thm1ran} and the fact that the randomized complexity is of order
$\e^{-2/(1+2r/d)}$. 

We summarize these results in the following theorem.
\begin{thm}\label{multintegra}
Consider the multivariate integration problem $\intt_{d,r}$ for
the class $F_{d,r}$. 
\begin{itemize}
\item Let $r=0$. 
 \begin{itemize} 
  \item In the quantum setting with deterministic queries, we have 
\begin{eqnarray*}
{\rm comp}^{{\rm que-std}}(\e,\intt_{d,0})\,&=&\,\infty,\\
{\rm comp}^{{\rm qub-std}}(\e,\intt_{d,0})\,&=&\,\infty.
\end{eqnarray*}
\item In the quantum setting with randomized queries, we have
\begin{eqnarray*}
{\rm comp}^{{\rm que-ran}}(\e,\intt_{d,0})\,&=&\,\Theta\left(\e^{-1}\right),\\
{\rm comp}^{{\rm qub-ran}}(\e,\intt_{d,0})\,&=&\,\Theta\left(\log\,\e^{-1}\right).
\end{eqnarray*}
\end{itemize}
\item Let $r\ge1$.
 \begin{itemize} 
  \item In the quantum setting with deterministic queries, we have 
\begin{eqnarray*}
{\rm comp}^{{\rm que-std}}(\e,\intt_{d,r})\,&=&\,\Theta\left(\e^{-1/(1+r/d)}\right),\\
{\rm comp}^{{\rm qub-std}}(\e,\intt_{d,r})\,&=&\,\Theta\left(\log\,\e^{-1}\right).
\end{eqnarray*}
\item In the quantum setting with randomized queries, we have
\begin{eqnarray*}
{\rm comp}^{{\rm que-ran}}(\e,\intt_{d,r})\,&=&\,\Theta\left(\e^{-1/(1+r/d)}\right),\\
{\rm comp}^{{\rm qub-ran}}(\e,\intt_{d,r})\,&=&\,\Theta\left(\log\,\e^{-1}\right).
\end{eqnarray*}
\end{itemize}
\end{itemize}
\end{thm}
\vskip 1pc
Hence, for $r=0$ we see a big difference between the two
settings for the multivariate integration problem,
whereas for $r\ge1$, the two settings lead to the same 
order of bit and qubit complexities. 

\section{Path Integration}
Path integration can be regarded as integration of functions
of infinitely many variables or, more formally, as integration over
some class of functions; for more information and references
see \cite{TW02}. Path integrals occur in quantum
physics, chemistry and mathematical finance. They are 
also the solutions of certain differential
equations and mathematical finance problems.

Here we consider a specific example of path integration studied in
\cite{TW02}. We take the space $X:=C([0,1])$ 
of continuous functions defined on $[0,1]$
with the norm $\|x\|=\max_{t\in [0,1]}|x(t)|$. The space $X$
is equipped with the classical Wiener measure $w$ for which
$$
\int_{X}x(t)\,w(dt)\,=\,0\ \ \forall\,t\in[0,1]\quad
\mbox{and}\quad
\int_{X}x(t)x(u)\,w(dx)\,=\,\min(t,u)\ \ \forall\,t,u\in[0,1].
$$
We consider the class $F$ of real
valued $w$-integrable functions $f:X\to\reals$ which are bounded
and satisfy a Lipschitz condition. 
More precisely, let the norm of $f$ be given by 
$\|f\|\,=\,\sup_{x\in X}|f(x)|$. Then the class $F$ is defined as
$$
F\,=\,\left\{\,f\,:\ \|f\|\,\le\,1,\
|f(x)-f(y)|\,\le\,\|x-y\|_{L_2([0,1])}\ \forall\,x,y\in X\ \right\}.
$$
Let $G=\reals$. The path integration $S:=\path$ is given by  
$$
{\path}(f)\,=\,\int_Xf(x)\,w(dx).
$$

We first consider the standard quantum setting. It was shown in
\cite{TW02} that we can compute 
an $\e$-approximation for path integrals from the class $F$ with
probability $\tfrac34$ using of order $\e^{-1}$ bit queries
and $\e^{-2}\log\,\e^{-1}$ qubits. 
The bound on the number of bit queries is sharp in the sense
that for any positive $\alpha$ it cannot be smaller that $\e^{1-\alpha}$ 
as $\e$ goes to zero. The sharpness of the number of qubits was not 
discussed. These bounds are obtained by reducing the path integration
problem to the summation problem which was solved by the 
Boolean summation algorithm. 

We consider the error defined by (\ref{randquant}) 
in the standard quantum setting. 
Due to the fact that the Boolean summation
algorithm enjoys optimality properties also in this setting, we
conclude that the same bounds as above also hold for the error
(\ref{randquant}). 
Furthermore, the bound on the number of qubits is sharp since the
worst case information 
complexity is of order $\e^{-c\,\e^{-2}}$ for some positive
$c$ as proved in \cite{curbera}.
Then Theorem \ref{thm1} yields that the number of qubits must 
be of order $\e^{-2}\log\,\e^{-1}$.

We now consider the quantum setting with randomized queries. 
It was proven in \cite{WW96} that the non-adaptive 
information complexity in the randomized classical setting is of the form
$$
{\rm comp}^{{\rm
    inf-ran}}(\e,\path)\,=\,\Theta\left(\e^{-2(1+o(1))}\right)
\quad\mbox{as} \ \e\,\to\,0.
$$
This and Theorem \ref{thm1ran} yields that the number of qubits is
at least of order $\log\,\e^{-1}$. 

We now show that in the quantum setting with randomized queries, 
we can solve the
path integration problem by using of order $\e^{-1}$ bit queries and
$\log\,\e^{-1}$ qubits. 
The space $X$ can be
embedded in the Hilbert space $L_2([0,1])$ for which the embedding 
${\rm Im}:X\to L_2([0,1])$, ${\rm Im}\,x=x$ for all $x\in X$, is a continuous
linear operator. Let $\nu=w\,{\rm Im}^{-1}$ be a zero mean Gaussian 
measure on $L_2([0,1])$. Then the covariance operator $C_{\nu}$ of the
measure $\nu$ has eigenpairs, $C_{\nu}\eta_i=\l_i\eta_i$, where   
$$
\eta_i(x)\,=\,\sqrt{2}\,\sin\left(\frac{2i-1}2\pi\,x\right),\qquad
\l_i\,=\,\frac4{\pi^2(2i-1)^2}.
$$
As in \cite{TW02} we first
approximate $\path(f)$ by
$$
\intt_d(f)\,=\,\int_{\reals^d}f_d(t)\,\mu_d(dt),
$$
where
$$
f_d(t)\,=\,f\left({\rm
    Im}^{-1}(t_1\eta_1+t_2\eta_2+\cdots+t_d\eta_d\right))
$$
and $\mu_d$ is the $d$ dimensional Gaussian measure with the mean zero
and variances $\l_i$. That is, its density function is of the form
$$
\frac1{(2\pi)^{d/2}\sqrt{\l_1\l_2\cdots\l_d}}\,\exp\left(-t_1^2/(2\l_1)-\cdots-
t_d^2/(2\l_d)\right).
$$
In \cite{TW02}, it is proved that for 
$d=\Theta(\e^{-2})$ we have  $|\path(f)-\intt_d(f)|\le\e/3$ for all
$f\in F$.

The integral $\intt_d(f)$ can be approximated by the Monte
Carlo
$$
\frac1n\sum_{j=1}^nf_d(t_j)
$$
with iid points $t_j$ distributed according to the measure $\mu_d$. 
Note that for $f\in F$, we have $|f_d(t_j)|\le 1$ and clearly the 
variance of $f_d$ is bounded by $1$. 
Therefore for $n=\lceil 9\e^{-2}\rceil$, the randomized error 
of approximating $\intt_d(f)$ is at most  $\e/3$, and 
the randomized error of approximating $\intt(f)$ is at most $2\e/3$. 
Finally, it is enough to use the Boolean summation algorithm to
approximate the last sum with the randomized error $\e/3$ which can be
done with of order $\e^{-1}$ bit queries and
$\log\,n=\Theta(\log\,\e^{-1})$ qubits. The randomized error of
approximation $\intt(f)$ is at most $\e$, as claimed. 

This and the previous lower bound on the number of queries yield 
that the randomized qubit complexity of path integration 
is of order $\log\,\e^{-1}$. For the randomized query complexity we have so far
an upper bound of order $\e^{-1}$. We can get a lower bound by
applying the same proof technique as in Theorem 3 of \cite{TW02}.
That is, the path integration problem is reduced to the real summation
problem for which we use a lower bound presented in Corollary
\ref{realran}. This yields that 
$$
\lim_{\e\to0}\e^{1-\alpha}\,{\rm comp}^{{\rm
    que-ran}}(\e,\path)\,=\,\infty
\quad\forall\,\alpha\in(0,1).
$$

We summarize these results in the following theorem.
\begin{thm}\label{path}
Consider path integration equipped with the Wiener measure for
the class $F$ of Lipschitz functions. 
\begin{itemize}
\item In the quantum setting with deterministic queries, we have 
\begin{eqnarray*}
{\rm comp}^{{\rm que-std}}(\e,\path)\,&=&\,\Theta\left(\e^{-1+o(1)}\right),\\
{\rm comp}^{{\rm qub-std}}(\e,\path)\,&=&\,\Theta\left(\e^{-2}\,\log\,\e^{-1}\right).
\end{eqnarray*}
\item In the quantum setting with randomized queries, we have
\begin{eqnarray*}
{\rm comp}^{{\rm que-ran}}(\e,\path)\,&=&\,\Theta\left(\e^{-1+o(1)}\right),\\
{\rm comp}^{{\rm qub-ran}}(\e,\path)\,&=&\,\Theta\left(\log\,\e^{-1}\right).
\end{eqnarray*}
\end{itemize}
\end{thm}
\vskip 1pc
The essence of this theorem is that for path integration 
we have an exponential improvement in the number of qubits 
in the quantum setting with randomized queries
whereas the number of queries remains 
roughly the same in both settings.
We stress that the optimal bounds for bit queries and qubits are
both attained by the same quantum algorithm based
on the Boolean summation algorithm.

\section{Appendix: Probabilistic Errors}

We briefly indicate what kind of results are possible if one studies
probabilistic errors in the randomized classical setting and in 
the quantum settings with
deterministic and randomized queries.

We begin with the randomized classical setting.
Instead of the randomized error (\ref{rand}) we now consider
the {\em probabilistic} error of the algorithm $A_n$
which is defined by the worst case performance with respect to
$f$ and the worst case performance with respect to $\omega$ modulo 
a set of measure $\delta$ for some (usually small) $\delta\in(0,1)$.
That is,
\begin{equation}\label{rand-prob}
\eran(A_n,\delta)\,=\,\sup_{f\in F}\
\inf_{B\subset\Omega,\,\rho(B)\le\delta}\ 
\sup_{\omega\in\Omega\setminus B}\|S(f)-A_n(f,\omega)\|.
\end{equation}
{}From Chebyshev's inequality we have 
\begin{equation}\label{cheb}
\eran(A_n,\delta)\,\le\,\frac1{\sqrt{\delta}}\,\eran(A_n).
\end{equation}
Better estimates with respect to $\delta$ are available under
additional assumptions on $S$. In any case, the dependence on $\delta$
is quite modest and everything depends on the randomized error
$\eran(A_n)$. This is probably why the probabilistic error
(\ref{rand-prob}) has not been 
as widely studied as the randomized 
error (\ref{rand}) for continuous problems on a classical computer.

The probabilistic error  yields the 
{\it (non-adaptive) information complexity} defined by
$$
{\rm comp}^{{\rm inf-ran}}(\e,\delta,S)\,=\,
\min\left\{\,n\,:\ \exists\, A_n^{{\rm nad}}\ \mbox{such that\ } e^{{\rm
        ran}}(A_n^{{\rm nad}},\delta)\,\le\,\e\,\right\}.
$$
Clearly, (\ref{cheb}) implies that
$$
{\rm comp}^{{\rm inf-ran}}(\e,\delta,S)\,\le\,
{\rm comp}^{{\rm inf-ran}}(\e\sqrt{\delta},S),
$$
\vskip 1pc
We now turn to the standard quantum setting. 
The usual way of defining the error in this setting 
is analogous to the probabilistic error.
That is, the {\em probabilistic} 
error of $A_{n,k}$ is defined as the smallest $\a$ for which 
$$
\|S(f)-A_{n,k}(f,j)\|\,\le\,\a
$$
holds with probability at most $1-\delta$ with respect to $j$ for 
every $f$ from $F$. Here, $\delta\in(0,1)$.
This definition can be formalized as follows.
For $f\in F$ and an arbitrary subset 
$J$ of $\{0,1,\dots,2^k-1\}$, let 
$\mu_f(J)=\sum_{j\in J}p_{f,j}$ be a measure of $J$. 
Then the probabilistic error of $A_{n,k}$ in the standard quantum setting is 
$$
\eqs(A_{n,k},\delta)\,=\,\sup_{f\in F}\ \min_{J:\,\mu_f(J)\le\delta}\
\ \max_{j\in \{0,1,\dots,2^k-1\}\setminus J}\|S(f)-A_{n,k}(f,j)\|.
$$

For some operators $S$, such as linear functionals, 
it is typical to take, say, $\delta=\tfrac14$, and obtain 
a quantum algorithm working with probability
$1-\delta$ by repeating several times the quantum algorithm working with 
$\delta=\tfrac14$ and by taking the median as the final result. If the 
number of repetitions is large enough we can 
boost probability of success to $1-\delta$. 
Details can be found in \cite{heinrich}.

The {\it probabilistic query complexity in the
standard quantum setting} is defined as 
\begin{equation}\label{qques-prob}
\cques(\e,\delta,S)\,=\,\min\left\{\,n\,:\ \exists\,A_{n,k}\ \mbox{such that\
  } \eqs(A_{n,k},\delta)\,\le\,\e\ \right\},
\end{equation}
and the {\it probabilistic qubit complexity in the standard quantum setting}
is defined as 
\begin{equation}\label{qqubs-prob}
\cqubs(\e,\delta,S)\,=\,\min\left\{\,k\,:\ \exists\,A_{n,k}\ \mbox{such that\
  } \eqs(A_{n,k},\delta)\,\le\,\e\ \right\}.
\end{equation}

\vskip .5pc
\noindent {\bf Example\,: Multivariate Integration (continued)}\
\newline
Assume that $r\ge1$. For $F=F_{d,r}$, it has been proven by Novak, see
\cite{N01}, that for $\delta=\tfrac14$
the minimal error of quantum algorithms $A_{n,k}$ is
of order $n^{-1-r/d}$ , and is achieved by an algorithm that
uses of order  $\log\,\e^{-1}$ qubits. 
The idea of the proof was to reduce the integration problem to 
real and then Boolean
summation and apply the the Boolean summation algorithm of
\cite{brassard}. This algorithm for the summation of $N$ terms with $n$ 
queries, $n\ll N$, has probabilistic error of order $n^{-1}$ which is 
optimal due to \cite{nayak}. 

This implies that the query complexity 
$\cques(\e,\tfrac14,\intt_{d,r})$ is of order $\e^{-1/(1+r/d)}$.
For $d$ much larger than $r$, we thus obtain roughly a quadratic speedup over
the randomized setting, and an exponential speedup over the worst case
setting. 

For arbitrary $\delta$, we can use roughly $\log\,\delta^{-1}$
repetitions of the algorithm used for $\delta=\tfrac14$ and take the
median of computed results as the final result, again see \cite{heinrich,novak}. This
implies that $\cques(\e,\delta,\intt_{d,r})$ is of order
$\e^{-1/(1+r/d)}\log\,\delta^{-1}$. 
 
For $r=0$, it is easy to see by the same argument which we used for
the randomized errors, that the integration problem cannot be solved
for the probabilistic error in the standard quantum setting.
\ \qed 

As before, Chebyshev's inequality yields
\begin{eqnarray*}
\cques(\e,\delta,S)\,&\le&\,\delta^{-1/2}\,\cques(\e,S)\\
\cqubs(\e,\delta,S)\,&\le&\,\delta^{-1/2}\,\cqubs(\e,S).
\end{eqnarray*}
Again the dependence on $\delta$ can be improved for some $S$.
Note, however, that even for general~$S$, the dependence on $\delta$
is quite weak.

We now show lower bounds on the probabilistic qubit complexity
in terms of the non-adaptive information complexities in the worst case
and randomized settings as well as in terms of the $\e$-entropy. 

\begin{thm}\label{thm1appprob}
\begin{eqnarray*}
\cqubs(\e,\delta,S)\,&\ge&\,\log\,\,\cw(2\e,S)\quad\forall\,\delta\in(0,\tfrac12),\\
\cqubs(\e,\delta,S)\,&\ge&\,\log\,\,\crr(\e,\delta,S),\\
\cqubs(\e,\delta,S)\,&\ge&\,{\rm Ent}(\e,S(F)).
\end{eqnarray*}
\end{thm}
\vskip 1pc
\noindent {\it Proof.\ } 
\newline
(1) To prove the first inequality,
we take a quantum algorithm $A_{n,k}$ 
which uses the minimal number of qubits $k=\cqubs(\e,\delta,S)$
with $\eqs(A_{n,k},\delta)\le\e$.  
We have 
\begin{eqnarray}
\e\,\ge\,\eqs(A_{n,k},\delta)&=&\sup_{f\in F}\,\min_{J:\,\mu_f(J)\le
  \delta}\ \ \max_{j\in\{0,1,\dots,2^k-1\}\setminus J}\, \|S(f)-A_{n,k}(f,j)\|\label{proof}\\&=&
\sup_{f\in F}\max_{j\in \{0,1,\dots,2^k-1\}\setminus J(f)}\|S(f)-A_{n,k}(f,j)\|,\nonumber
\end{eqnarray}
where $J(f)$ is a subset of $\{0,1,\dots,2^k-1\}$,
$\mu_f(J(f))\le\delta$,  for which the
corresponding minimum is attained. Such a set exists since we have
finitely many such subsets, however, $J(f)$ is not necessarily unique. 
Let 
$$
M(f)\,=\,\{\,0,1,\dots,2^k-1\,\}\,\setminus \,J(f).
$$
Clearly, $\mu_f(M(f))\ge1-\delta$. 

For an arbitrary $f\in F$, we take two functions $f_1$ and $f_2$ such
that $N(f_1)=N(f_2)=N(f)$. Note that the measures $\mu_{f_1}$ and
$\mu_{f_2}$ are the same. 
For $\delta<\tfrac12$, there exists an index $j^*$ which belongs to  
$M(f_1)\cap M(f_2)$. Indeed, otherwise the sets $M(f_1)$ and $M(f_2)$
would be disjoint and  
\begin{eqnarray*}
1\,\ge\, \mu_f(M(f_1)\cup M(f_2))\,&=&\,
\mu_f(M(f_1))+\mu_f(M(f_2))\\\,&=&\,
\mu_{f_1}(M(f_1))+\mu_{f_2}(M(f_2))\,\ge\,2(1-\delta)\,> 1
\end{eqnarray*}
is a contradiction. For this index $j^*$, we have
$a=A_{n,k}(f_1,j^*)=A_{n,k}(f_2,j^*)$. From (\ref{proof}) we get 
$$
\e\ge\tfrac12\left( \|S(f_1)-a\|+\|S(f_2)-a\|\right)\,\ge \tfrac12\|S(f_1)-S(f_2)\|.
$$
Hence,
$$
\,\sup_{f\in F}\sup_{f_1,f_2\in
  F,\,N(f_1)=N(f_2)=N(f)}\|S(f_1)-S(f_2)\|\,\le\,2\e,
$$
and the rest of the proof is the same as in the proof of Theorem
\ref{thm1}. 

(2) To prove the second inequality, we compare the qubit complexity to the randomized
non-adaptive information complexity. Observe that any quantum
algorithm $A_{n,k}$ may be regarded as a randomized algorithm which uses
non-adaptive deterministic information of cardinality at most $2^k$
with randomized elements $\omega\in\{0,1,\dots,2^k-1\}$ taking values
$j$ with probability $p_{f,j}$. Furthermore,
the probabilistic error of $A_{n,k}$ is exactly the same as the error 
in the probabilistic randomized setting. 
Therefore, $\eran(A_{n,k},\delta)\le\e$ can hold only if the cardinality
of $A_{n,k}$ is $\crr(\e,\delta,S)$. This means that
$2^k\ge\crr(\e,\delta,S)$, as claimed.

(3) To prove the third inequality, observe that we now have for any
$f\in F$, 
$$
\max_{j\in M(f)}\|S(f)-\phi(j)\|\,\le\,\e,
$$
where the subset $M(f)$ of $\{0,1,\dots,2^k-1\}$ 
is defined as above. Hence, 
$$
\min_{j=0,1,\dots,2^k-1}\|S(f)-\phi(j)\|\,\le\,\e,
$$
and the rest is as in the proof of Theorem \ref{thm2}.\ \qed
\vskip 1pc
We now turn to the quantum setting with randomized queries.
Instead of (\ref{randquant}), we consider the 
{\em probabilistic} error of $A_{n,k}$ which is defined as
the smallest $\a$ for which
$$
\|S(f)-A_{n,k}(f,\omega,j)\|\,\le\,\a
$$
holds with probability at least $1-\delta$ with respect to $j$ and 
$\omega$ for all $f$ from $F$. More
precisely, as before, for $J\in\{0,1,\dots,2^{k}-1\}$ we define 
the measure of~$J$ by $\mu_{f,\omega}(J)=\sum_{j\in J}p_{f,j,\omega}$. 
Then the probabilistic error of $A_{n,k}$ in the quantum setting 
with randomized queries is 
$$
\eqr(A_{n,k},\delta)\,=\,\sup_{f\in F}\ \inf_{B\in\Omega,\,\rho(B)\le\delta} 
\sup_{\omega\in  \Omega\setminus B}
\min_{J:\,\mu_{f,\omega}(J)\le\delta}\ \max_{j\in
  \{0,1,\dots,2^{k}-1\}\setminus J}\ 
\|S(f)-A_{n,k}(f,\omega,j)\|.
$$
Observe that if we choose $A_{n,k}$ independently of $\omega$ then, modulo
measurement, everything will be deterministic
and the last definition coincides with the probabilistic error
in the standard quantum setting.

The probabilistic query/qubit complexity in the quantum setting 
with randomized queries are
defined, analogously as in the standard quantum setting, by minimizing
the number of queries/qubits needed to find a quantum
algorithm whose probabilistic error is at most
$\e$.  That is, 
the {\it probabilistic query complexity in the quantum
  setting with randomized queries} 
is defined by 
$$
\cquer(\e,\delta,S)\,=\,\min\left\{\,n\,:\ \exists\,A_{n,k}\ \mbox{such that\
  } \eqr(A_{n,k},\delta)\,\le\,\e\ \right\},
$$
and the {\it probabilistic qubit complexity in the quantum
  setting with randomized queries} is defined by 
$$
\cqubr(\e,S,\delta)\,=\,\min\left\{\,k\,:\ \exists\,A_{n,k}\ \mbox{such that\
  } \eqr(A_{n,k},\delta)\,\le\,\e\ \right\}.
$$

We now show that Chebyshev's inequality implies
\begin{equation}\label{cheb2}
\eqr(A_{n,k},\delta)\,\le\,\delta^{-1}\ \eqr(A_{n,k}).
\end{equation}
Indeed, let 
$$
E(f,\omega)\,=\,
\sum_{j=0}^{2^{k}-1}p_{f,j,\omega}\|S(f)-A_{n,k}(f,\omega,j)\|^2.
$$
Hence,
$$
\eqr(A_{n,k})^2\,=\,\sup_{f\in
  F}\int_{\Omega}E(f,\omega)\,\rho(d\omega).
$$
Define the sets $B(f)$ and $J(f,\omega)$ by
\begin{eqnarray*}
\Omega\setminus B(f)\,&=&\,\left\{\omega:\
  E(f,\omega)\,\le\,\delta^{-1}\eqr(A_{n,k})^2\,\right\},\\
\left\{0,1,\dots,2^{k_{\omega}}-1\right\}\setminus J(f,\omega)\,&=&\,\left\{
j:\, \|S(f)-A_{n,k}\|^2\,\le\,\delta^{-1}E(f,\omega)\,\right\}.
\end{eqnarray*}
Chebyshev's inequality tells us that $\rho(B(f))\le\delta$ and
$\mu_{f,\omega}(J(f,\omega))\le\delta$.
Then
$$
\eqr(A_{n,k},\delta)^2\,\le\,\sup_{f\in F}\ \sup_{\omega\in\Omega\setminus
  B(f)}\ \delta^{-1}\ E(f,\omega)\,\le
\delta^{-2}\ \eqr(A_{n,k})^2,
$$
as claimed.

The probabilistic and randomized query and qubit complexities are related.
{}From (\ref{cheb2}) we have
\begin{eqnarray*}
\cquer(\e,\delta,S)\,&\le&\,\cquer(\e\delta,S),\\
\cqubr(\e,\delta,S)\,&\le&\,\cqubr(\e\delta,S).
\end{eqnarray*}

It is also easy to see to check that
\begin{eqnarray*}
\cqubr(\e,S)\,&\ge&\,\log\,{\rm comp}^{{\rm inf-ran}}(\e,S),\\
\cqubr(\e,\delta, S)\,&\ge&\,\log\,{\rm comp}^{{\rm inf-ran}}(\e,\delta,S).
\end{eqnarray*}
\section{Acknowledgment}
 I am grateful for many discussions and valuable comments from
 S. Heinrich, M. Kwas, E. Novak, A. Papageorgiou and J. F. Traub.

\end{document}